\documentclass[proof]{pasj00}
\SetRunningHead{Y. MIYAMOTO et al.}{Evolution of Molecular Clouds by Shear Motion}
\Received{2013 April 22}
\Accepted{2013 November 15}
\Published{}
\begin{document}
\title{Influence of Shear Motion on Evolution of\\ Molecular Clouds in the Spiral Galaxy M51}
\author{Yusuke \textsc{Miyamoto}\altaffilmark{1}, Naomasa \textsc{Nakai}}
\affil{ Division of Physics, Faculty of Pure and Applied Sciences, University of Tsukuba, Ten-noudai, Tsukuba, Ibaraki 305-8571}
\email{miya@mx.ibaraki.ac.jp}
\and
\author{Nario \textsc{Kuno} \altaffilmark{}}
\affil{Nobeyama Radio Observatory, Minamimaki, Minamisaku, Nagano 384-1305}
\footnotetext[1]{Present Adress: Center for Astronomy, Ibaraki University, 2-1-1 Bunkyo, Mito, Ibaraki 310-8512}

\maketitle

\begin{abstract}
We have investigated the dynamics of the molecular gas and the evolution of GMAs in the spiral galaxy M51 with the NRO 45-m telescope.
The velocity components of the molecular gas perpendicular and parallel to the spiral arms are derived at each spiral phase from the distribution of the line-of-sight velocity of the CO gas. 
In addition, the shear motion in the galactic disk is determined from the velocity vectors at each spiral phase. 
It is revealed that the distributions of the shear strength and of GMAs are anti-correlated. 
GMAs exist only in the area of the weak shear strength and further on the upstream side of the high shear strength.
GMAs and most of GMCs exist in the regions where the shear critical surface density is smaller than the gravitational critical surface density, indicating that they can stably grow by self-gravity and the collisional agglomeration of small clouds without being destroyed by shear motion.
These indicate that the shear motion is an important factor in evolution of GMCs and GMAs.
\end{abstract}

\section{Introduction}
\begin{table}
	\begin{center}
	\caption{The parameters of M51}
		\begin{tabular}{ccc}
		\hline
		\hline
	Center position 					& $R.A.$(J2000.0) : \timeform{13h 29m 52s.711} 		& \citet{turner1994} \\
					 			&$ Decl.$(J2000.0) : \timeform{47D 11' 42''.61} 			&\\
	Distance 		 				& 9.6 Mpc 									& \citet{sandage1974}\\
	Morphological type 				& SAbc 										& \citet{devaucouleurs1991}\\
	Systemic Velocity (LSR)			& $469\pm4$~km~s$^{-1}$ 						& This paper\\
	Position angle of the major axis 	& $-\timeform{9D}\pm\timeform{6D}$ 				& This paper\\
	Inclination angle 				& \timeform{22D}$\pm\timeform{3D}$				& This paper\\ 
		\hline
		\hline
		\end{tabular}
	\end{center}
\label{tab:parameter}
\end{table}

The evolution of molecular clouds is one of the keys to understand 
star formation in a galaxy and hence evolution of the galaxy. 
Massive stars are formed in GMCs (Giant Molecular Clouds), 
and GMCs constitute GMAs (Giant Molecular Associations), 
where typical sizes of a GMC and a GMA are a few 10 pc \citep{sanders1985}, 
and a few 100~pc \citep{vogel1988}, respectively. 
Spiral arms are a key to understand evolution of GMCs. 
Observationally the molecular arms are traced by CO lines with small offset relative to the optical arms identified by H$\alpha$ (e.g., \cite{vogel1988}). 
A density wave \citep{lin1964} generates a galactic shock in the spiral potential \citep{fujimoto1968}
which explains the offset of the optical arms from the molecular arms. 
Massive star formation could be induced by the shock and cloud-cloud collisions in the molecular arms \citep{roberts1990}, 
so that the optical arms are located at the slightly downstream side of the molecular arms. 
GMCs can be destroyed by photodissociation owing to UV radiation from massive stars 
and shocks caused by supernova explosions  
(e.g., \cite{mckee1977}, \cite{allen1986}, \cite{seta1998}). 
The destroyed gas would be in the atomic phase rather than the molecular phase in the downstream 
and interarm regions as far as the conventional density wave and galactic shock theories hold. 
However, detections of GMCs and GMAs have been reported in the interarm regions recently. 
In the grand design spiral galaxy M51, \citet{koda2009} showed that GMCs (M $\sim$ $10^5$ - $10^6$~\MO) exist not only in the arms but also in the interarm regions. 
They also found GMAs ($\leq10^7$~\MO) in the arms and the remnants of fragmented GMAs in the interarms. 
From these results they suggested that GMAs were fragmented into GMCs by kinematic shear caused by abrupt change in velocity. 
\citet{muraoka2009} found that molecular clouds in the interarms of M83 were not necessarily virialized, 
while molecular clouds in the arms were virialized, 
and suggested that the non-virialized molecular clouds in the interarms might be formed by the shear motion. 
Strong shear may generate turbulence in the clouds that suppresses star formation. 
However, the relation between molecular clouds and the kinetic shear motion in the clouds is still speculation, 
because the kinetic shear in the scale of GMCs and/or GMAs has not been directly measured in a galaxy.
We have to know the strength and length of the shear observationally. 

The nearby galaxy  M51 at the distance of 9.6~Mpc \citep{sandage1974} has a nearly face-on disk (inclination angle $i=\timeform{20D}$: \cite{tullyb1974}) with the grand design two-fold spiral structure. 
Although the galaxy is interacting with its companion galaxy NGC 5195, 
the inner disk is less affected (see section \ref{subsec:spiral}).
Since CO emission is fairly strong throughout the galactic disk, 
the distribution and kinematics of molecular gas can be easily measured.
Thus M51 is an ideal target to obtain the shear motion in the galactic disk. 
Using the data of $^{12}$CO($J=1$$-$$0$) newly mapped with the Nobeyama 45-m telescope, 
we determined the kinetic shear in the whole disk of the galaxy from velocity vectors of molecular gas obtained by applying the method of \citet{kuno1997}.
In this paper, we report the results and discuss the relation between the strength of the shear and formation and destruction of GMCs and GMAs.
The basic parameters of M51 adopted in this paper are summarized in table~\ref{tab:parameter}.
Velocities used here are in the radio definition and with respect to the local standard of rest (LSR).
The LSR velocity, $V_{\rm LSR}$, is converted from the heliocentric velocity, $V_{\rm helio}$, by using $V_{\rm LSR} = V_{\rm helio} + 11.7$~km~s$^{-1}$ for this galaxy.

\section{Observations}
\label{chap:obs}
Observations of $^1$$^2$CO($J=1$$-$$0$) emission were carried out between 2004 December and 2007 March using the 45-m telescope of the Nobeyama Radio Observatory (NRO).
The antenna was equipped with the $5\times5$-beams SIS heterodyne receiver array (BEARS), 
which could measure twenty five spectra of CO simultaneously (\cite{sunada2000}). 
The beam separation of BEARS was 41.2$''$, 
and the full half power beam width (HPBW) was $16''$ at 115~GHz, 
corresponding to 745~pc at the distance of 9.6~Mpc (\cite{sandage1974}). 
The receiver backends were the 1024-ch digital spectrometers (\cite{sorai2000}). 
The total bandwidth and frequency resolution of the spectrometers were 512~MHz and 605~kHz, 
which correspond to 1330~km~s$^{-1}$ and 1.57~km~s$^{-1}$, respectively. 
The line intensity was calibrated by the chopper wheel method, 
yielding an antenna temperature, $T_{\rm A}^{\ast}$, corrected for both atmospheric and antenna ohmic losses (\cite{ulich1976}). 
In this paper we use the main beam brightness temperature $T_{\rm mb}\equiv T_{\rm A}^{\ast}/\eta_{\rm mb}$, using the main beam efficiency of the antenna, $\eta_{\rm mb} =$ 0.32 to 0.39. 
The observations were made in the position-switching mode with an integration time of 20~s per scan and OFF-positions offset by $\pm8'$ in the direction of the azimuth. 
The intensity was calibrated every 10 sequences of ON(source)-OFF(sky) positions. 
The telescope pointing was checked every 45~min - 1~hr by observing SiO maser emission of the late-type star R CVn. 
The pointing error was mostly less than $\rm5''$, but $\rm2''$ in the center region of the galaxy. 
The system noise temperatures during the observations were 600-1100~K (SSB) in $T_{\rm A}^{\ast}$ at observing elevations. 
We observed 2760~points in the region of about $9' \times 10'$ (but no point in north-west of the map), 
which contained the whole optical disks of M51 and NGC 5195, and the bridge between them.  
Observed grid points were located parallel to the major (X) and the minor (Y) axes of the galactic disk whose position angle was $-\timeform{10D}$ \citep{tullyb1974}. 
The transformation from the (${X, Y}$) coordinate to ({\it dR.A., dDecl.}) is given by  
$dR.A. = 0.9848X - 0.1736Y$, 
$dDecl. = 0.1736X + 0.9848Y$, 
adopting the position angle of the disk to be $\theta_{\rm PA} = -\timeform{10D}$. 
The interval of the grid points were 10$''$.

\section{Results}
\label{chap:res}
\subsection{CO Spectra}
\label{sec:spe}
\begin{figure}[tp]
\begin{center}
 \FigureFile(120mm,100mm){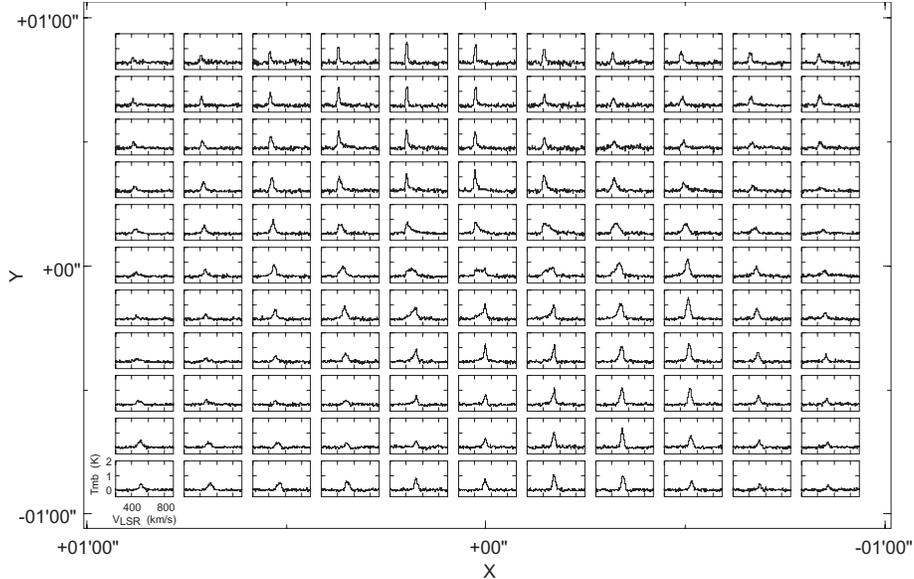}
\end{center}
\caption{
Measured spectra of $^{12}$CO(1$-$0) emission in the central region of M51, parallel to the minor (X) ($\theta_{\rm PA}=\timeform{80D}$) and major (Y) ($\theta_{\rm PA}=\timeform{-10D}$) axes of the galactic disk. 
For each spectrum the abscissa is the LSR velocity ($V_{\rm LSR} = 200$ - $900$~km~s$^{-1}$) and the ordinate is the main-beam brightness temperature ($T_{\rm mb} = -0.5$~K-2.0~K). The grid spacing between the spectra is $10''$.}
\label{fig:m51cent}
\end{figure}
\begin{figure}[htbp]
\begin{center}
 \FigureFile(120mm,100mm){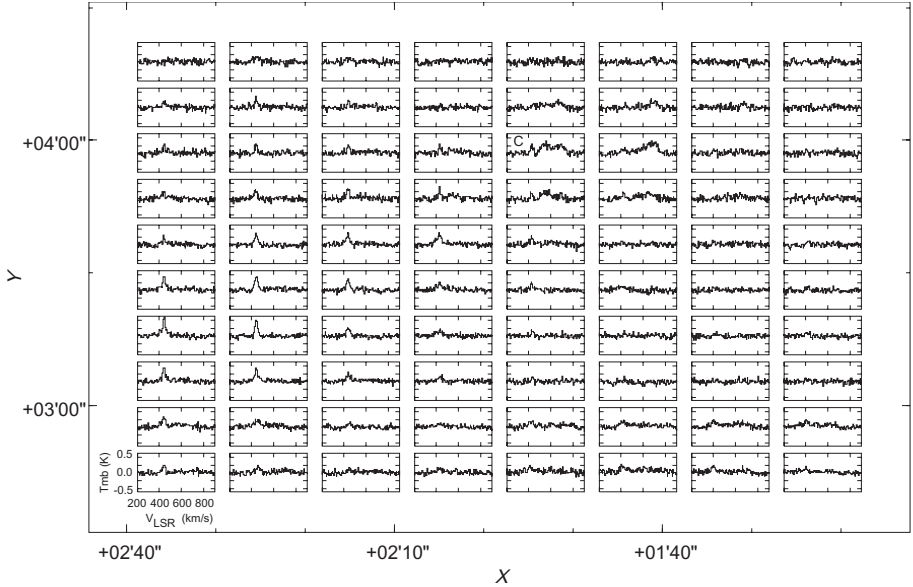}
 \end{center}
\caption{Same as figure~\ref{fig:m51cent} but in the vicinity of the companion galaxy NGC 5195.
The ordinate range is $T_{\rm mb} = -0.5$~K-0.5~K.
The spectrum with a character C contains the center of the galaxy,
{\it R.A.}(J2000.0) $= 13^{\rm h} 29^{\rm m} 59^{\rm s}.54$ and {\it Decl.}(J2000.0) $= \timeform{47D} 15' 58.0 '' $ \citep{hummel1987}, but being shifted by $\Delta R.A. = 0.1'', \Delta Decl. = -2.3''$.
Note a narrow component at $V_{\rm LSR} = 400$ - $450$~km~s$^{-1}$
and a wide component at $V_{\rm LSR} = 450$ - $750$~km~s$^{-1}$.
}
\label{fig:5195}
 \end{figure}

All the measured $\rm^{12}CO(1$$-$$0)$ spectra have been smoothed to a velocity resolution of 5~km~s$^{-1}$ to reduce noise.
One $\sigma$ of the $I_{\rm CO}$, i.e., ${\rm \Delta} I_{\rm rms}$, is expressed by $ {\rm  \Delta} I_{\rm rms} = {\rm \Delta} T_{\rm rms} \sqrt{\mathstrut {\rm \Delta} V_{\rm FWHM} {\rm \Delta} V_{\rm c}}$, where ${\rm \Delta} V_{\rm c}$ is the smoothed velocity resolution. 
Figure~\ref{fig:m51cent} shows the spectra in the central region of M51.
The maximum of the main beam brightness temperature is $T_{\rm mb} = 1.6$~K.
In the disk of M51, the brightness temperature in the spiral arms 
and the interarms are $T_{\rm mb} \geq$ 0.5 - 1~K and $\approx$ 0.2 - 0.4~K, respectively, 
which are consistent with the previous values measured with the 45-m telescope \citep{nakai1994}. 
The rms noise level, $\Delta T_{\rm rms}$, is less than 0.1 K in the disks of M51 and NGC 5195, 
and is larger at the edges of the mapping area.
The velocity width of the CO line in the M51 disk is typically $ \Delta V_{\rm FWHM}  \approx20$~km~s$^{-1}$, 
but gradually decreasing with the galactic radius from $\Delta V_{\rm FWHM} \rm \approx 45$~km~s$^{-1}$ at 
$r=40''$ to $\rm \approx 6$~km~s$^{-1}$ at $r =140''$, where $r$ is the galactrocentric distance.

Figure~\ref{fig:5195} shows the spectra toward the disk of NGC 5195, 
the brightness temperature is low ($ T_{\rm mb} \approx 0.3$-0.4~K), 
but the velocity width is very wide ($\Delta V_{\rm FWHM}\rm \approx350$~km~s$^{-1}$).
In more detail, the spectra show two velocity compornents of CO emission,
a narrow component at $V_{\rm LSR} \approx \rm 400$-450~km~s$^{-1}$ 
and a very wide component at $V_{\rm LSR} \approx \rm 450$-750~km~s$^{-1}$ in the vicinity of ($X,Y$)=(1$'$ 45$''$, 4$'$ 00$''$). 
The two distinct components have been pointed out in the earlier study \citep{sage1989}. 
Figure~\ref{fig:n5195-overlay} shows the distributions of the narrow and wide components separately.
Since the distribution of the narrow component is connected to a spiral arm of M51 (section~\ref{sec:dist}), 
the component arises from the overlying spiral arm of M51,
while the wide component, which includes the systemic velocity ($V_{\rm LSR} = 642$~km~s$^{-1}$, \cite{sage1989})
of NGC 5195, is associated with NGC 5195.

\subsection{ Distribution of the CO Intensity}
\label{sec:dist}
\begin{figure}[tp]
\begin{center}
 \FigureFile(80mm,100mm){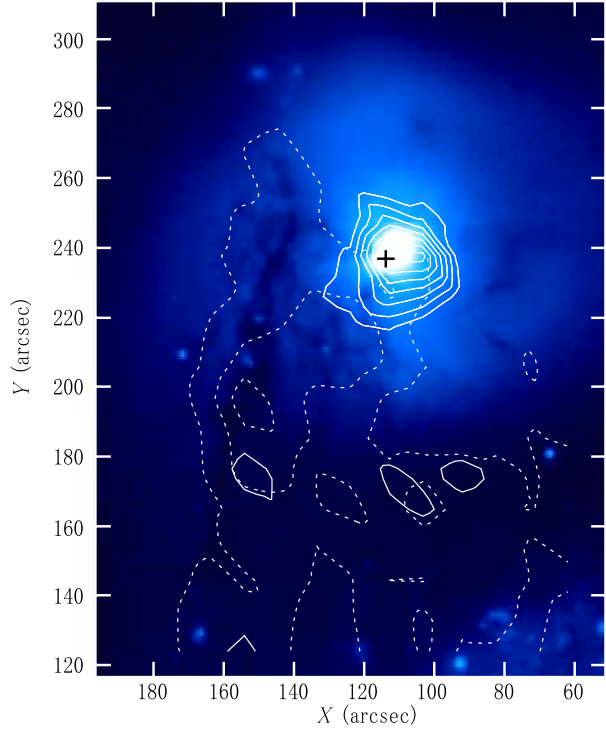}
 \end{center}
\caption{\atom{C}{}{12}O (1$-$0) intensity maps of the narrow component integrated at $V_{\rm LSR} = 400$-450~km~s$^{-1}$ (dotted contours) and of the wide component integrated at $V_{\rm LSR} = 450$-750~km~s$^{-1}$ (solid contours) overlaid on the R-band image (NASA$/$IPAC Extragalactic Database (NED); \cite{hoopes2001}). 
The first contours and contour intervals are 5~K~km~s$^{-1}$ and 5~K~km~s$^{-1}$
for the narrow component (dotted lines) 
and 10~K~km~s$^{-1}$ and 5~K~km~s$^{-1}$  
for the wide component (solid lines), respectively.
The narrow and wide components arise from an overlying spiral arms of M51 and from molecular gas inside NGC 5195, respectively.
A cross denotes the center of NGC 5195.}
\label{fig:n5195-overlay}
 \end{figure}

\begin{figure}[tp]
\begin{tabular}{cc}
 \begin{minipage}{0.5\hsize}
  \par
  {\textbf{(a)}}
 \begin{center}
 \FigureFile(80mm,50mm){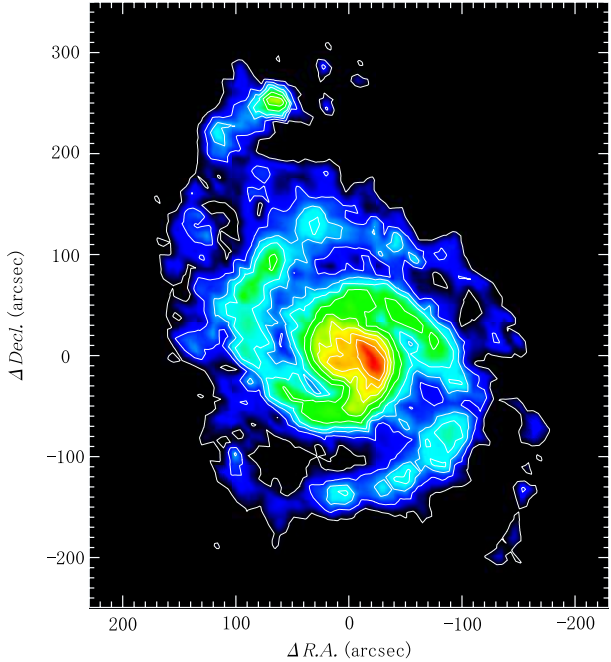}
   \end{center}
   \end{minipage}
 \begin{minipage}{0.5\hsize}
\par
{\textbf{(b)}}
 \begin{center}
 \FigureFile(80mm,50mm){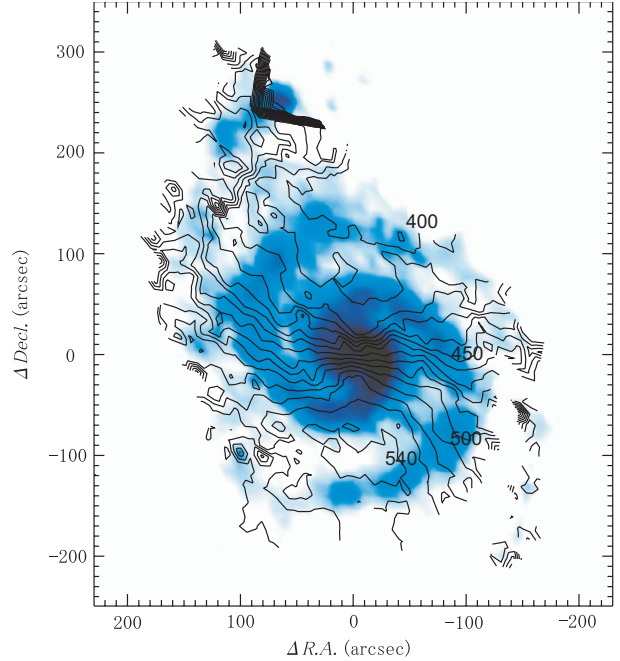}
 \end{center}
 \end{minipage}\\
 \end{tabular}
\caption{\textbf{(a)} The map of the $^{12}$CO(1$-$0) integrated intensity, $I_{\rm CO}(1$$-$$0)\equiv\int T_{\rm mb}dv$ [K~km~s$^{-1}$], in M51 and its companion galaxy NGC 5195 in the northeast. 
Contours are from 5, 10, 15, 20 to 70~K~km~s$^{-1}$ in steps of 10~K~km~s$^{-1}$.
The origin of the coordinates is the center of M51 (table~\ref{tab:parameter}). 
The image  resolution is 16$''$.\\
\textbf{(b)} The $^{12}$CO(1$-$0) velocity field map (solid contours) derived from intensity-weighted mean velocities overlaid with the distribution of $I_{\rm CO}$ (colour). The velocity is in km~s$^{-1}$ with respect to LSR and the radio definition.
}
 \label{fig:map}
\end{figure}
Figure~\ref{fig:map}a shows the distribution of the CO integrated intensity, 
$I_{\rm CO} \equiv \int T_{\rm mb}dv$ (K~km~s$^{-1})$.
The distribution of $I_{\rm CO}$ shows two spiral arms as seen in the previous maps (\cite{rand1990}, \cite{garcia1993a}, \cite{nakai1994}, \cite{aalto1999}, \cite{helfer2003}, \cite{shetty2007}, \cite{schuster2007}, \cite{koda2009}, 2011, \cite{vlahakis2013}).
One is extending to the companion NGC 5195 and another, the counterpart arm, 
is extending toward the southwest. 
The $\rm CO(1$$-$$0)$ distribution is in agreement with interferometric observations of $\rm CO(1$$-$$0)$ (\cite{shetty2007}, \cite{koda2011})  
and analogous to the distribution of CO(2$-$1) \citep{schuster2007} except for the vicinity of NGC 5195.
The curvature of the $\rm CO(1$$-$$0)$ distribution at  the connection between M51 and NGC 5195 in figure~\ref{fig:n5195-overlay} and figure~\ref{fig:map}a
is steeper than that of CO(2$-$1) in \citet{schuster2007}. 
The difference between those distributions can be attributed to the narrowness of the integration range in CO(2$-$1),
$V_{\rm LSR} = 350$ - 600~km~s$^{-1}$, 
despite of the velocity components around NGC 5195 extend out to 750~km~s$^{-1}$.
In addition, the distribution of CO(3$-$2) \citep{vlahakis2013}, which is similar to that of CO(2$-$1) rather than CO(1$-$0),  
is also explained by the limited integration range ($V_{\rm LSR} \approx $ 380 - 580~km~s$^{-1}$). 
The distribution of $\rm CO(1$$-$$0)$ in figure~\ref{fig:n5195-overlay} corresponds with the dust lane in the R-band image \citep{hoopes2001}.

\subsection{Basic Parameters of M51}
\label{subsec:para}
\subsubsection{Dynamical Center and Systemic Velocity}
\label{subsec:vsys}

\begin{figure}[tp]
\begin{center}
 \FigureFile(100mm,80mm){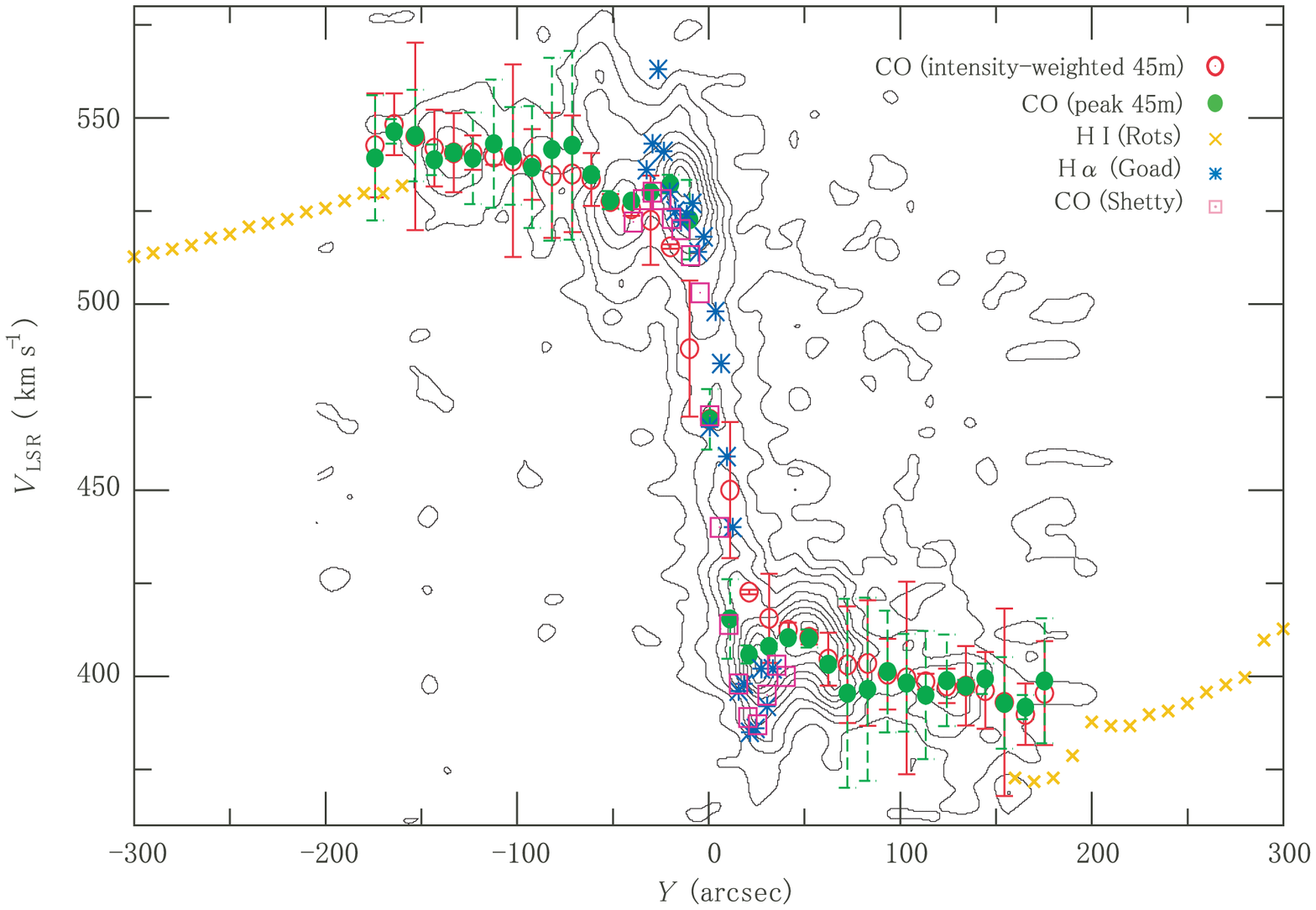}
 \end{center}
\caption{Position-velocity diagram along the major axis (Y) of M51. 
The lowest contour and the contour interval of the main beam brightness temperature are 0.15~K. Other data of velocities are also
plotted; H\emissiontype{I} observed with the VLA \citep{rots1990}, H${\alpha}$ with the Kitt Peak 4-m
telescope \citep{goad1979} and CO(1-0) with the BIMA \citep{shetty2007}.}
\label{fig:V_lsr}
\end{figure}
 Figure~\ref{fig:V_lsr} shows the position-velocity diagram of CO (contours) along the major axis  of M51
 with the position angle of $\theta_{\rm PA}=\timeform{-10D}$ \citep{tullyb1974}.
The dynamical center and the systemic velocity were derived from the position-velocity diagram 
by averaging the CO intensity-weighted velocity at $|R| < 150''$, 
assuming that the velocity field was symmetrical to the dynamical center.
The  resultant systemic velocity was $ V_{\rm sys} = 469\pm4$~km~s$^{-1}$
which was consistent with the result of \citet{tullyb1974}, $ V_{\rm sys} = 475 \pm 3$~km~s$^{-1}$ in radio definition and with respect to LSR.
As shown in the next subsection, our derived position angle is $\theta_{\rm PA}=\timeform{-9D}$
which is slightly different from $\theta_{\rm PA}=\timeform{-10D}$.
Even if we use $\theta_{\rm PA}=\timeform{-9D}$, the difference of $V_{\rm sys}$ is less than 1 ~km~s$^{-1}$.  Therefore we adopt $V_{\rm sys}=469$~km~s$^{-1}$ hereafter.

The dynamical center  which showed the systemic velocity was $\Delta Y = + 2'' \pm 2''$ away from the adopted center of the $\rm5$-GHz continuum peak \citep{turner1994}. 
Since the position is consistent with that of the continuum peak within the error and 
the apparent difference of $\Delta Y = +2''$ is much smaller than our angular resolution ($16''$),
we use the position of the radio continuum peak whose accuracy was \timeform{0''.1}-\timeform{0''.5},
as the center of M51 for analyses in discussion.


\subsubsection{Position Angle}
\label{subsec:pa}

\begin{figure}[tp]
\begin{center}
 \FigureFile(100mm,80mm) {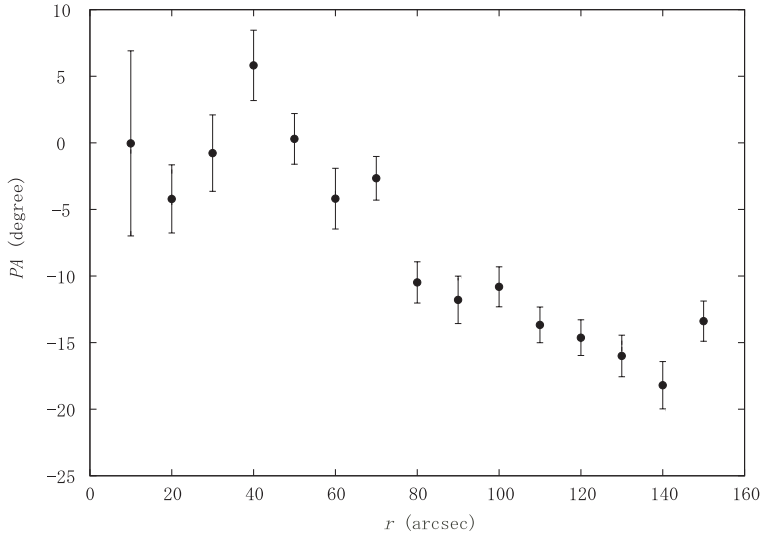}
\caption{Radial variation of the position angle of the major axis of the M51 disk.
The position angles were derived by fitting observed velocity data (figure \ref{fig:map}b) 
in the every 10$''$ circular rings from $r =10''$ to 150$''$ assuming circular rotation.
}
\label{fig:pa}
\end{center}
\end{figure}

The position angle of the major axis of the M51 disk has been determined from the kinematics of $\rm H{\alpha}$ and CO (\cite{tullyb1974}, \cite{kuno1997}, \cite{shetty2007}).
We determined it in a similar way to \citet{kuno1997}, using our new data.
The galactic disk was divided into annuli with a width of $\Delta r=10''$ centered on the galactic center, 
and the observed radial velocities at observed points ($r$, $\theta_{\rm obs}$) in each annulus ($r=$ distance from the center, $\theta_{\rm obs}=$ position angle), 
$V_{\rm rad} \left[=V_{\rm sys} -V_{\rm gal,o} \cos(\theta_{\rm obs} -\theta_{\rm PA})\right]$, were fitted with two parameters, 
the observed rotation velocity, $V_{\rm gal,o}$, and the position angle of the galaxy, $\theta_{\rm PA}$, where $V_{\rm sys}$ is 469~km~s$^{-1}$. 
Figure~\ref{fig:pa} shows the result of $\theta_{\rm PA}$ derived by this method.
The position angle declines with the galactic radius from $\theta_{\rm PA} \approx \timeform{5D}$ at $r = 40''$ to $\approx - \timeform{20D}$ at $r = 140''$.
The mean values of $\theta_{\rm PA}$ in the range of $r = 10''$ - $150''$ 
($r = 40''$ - $70''$, $70''$ - $110''$ and $110''$ - $140''$) are $-\timeform{9D}\pm \timeform{6D}$ 
($-\timeform{3D}$, $-\timeform{10D}$ and $-\timeform{15D}$, respectively), 
which is consistent with the results of Tully ($\theta_{\rm PA} = -\timeform{10D}\pm\timeform{3D}$, 1974) and Kuno \& Nakai ($\theta_{\rm PA} = -\timeform{8.4D}$, 1997).

It is to be noted that the position angle derived by this method is biased from the true value in the counter-clockwise direction due to the non-circular motion of the molecular gas influenced by the spiral potential.
The spiral arms cross the major axis in the region of $r = 40''$ - $70''$ (see figure \ref{fig:V_lsr}),  
where the deviation from the rotation of the disk ($V_{\rm gal,o}\sim 70$~km~s$^{-1}$)  
is of $\Delta V_{\rm gal,o}\sim 10$~km~s$^{-1}$.
This offset induces the bias of the position angle $\theta_{\rm PA}$ of $\Delta\theta_{\rm PA}\sim \timeform{10D}$ in the counter-clockwise direction.
However, the systematic variation of the position angle at $r=10''$ - $150''$ in figure \ref{fig:pa} cannot be explained by the effect of only the non-circular motion but of the warped disk which may be caused by interaction with the companion NGC 5195 (e.g., \cite{shetty2007}).

\subsubsection{Inclination Angle}
\label{subsec:incl}
We estimated the inclination angle ($i$) of the M51 disk by using 
the baryonic Tully-Fisher relation (\cite{mcgaugh2005}, \cite{shetty2007}), where $i=\timeform{90D}$ is edge-on.
\citet{mcgaugh2005} showed the relation of $M_{\rm b} = 50 V^4_{\rm rot}$ between the total baryonic mass of a galaxy, $M_{\rm b} (= M_{\rm star} + M_{\rm gas})$, and the rotation velocity of it, $V_{\rm rot}=V_{\rm gal,o}/\sin i$.
We used the baryonic masses in M51 of $M_{\rm star}=5.2^{+1.4}_{-1.1}\times 10^{10}$ \MO (HyperLeda database, \cite{bell2003}) 
and $M_{\rm gas} =1.36\times( M_{\rm H_2}+M_{\rm H\emissiontype{I}}+M_{\rm H\emissiontype{II}})$ including helium, 
where $M_{\rm H_2}=(5.2\pm0.7) \times 10^9$ \MO  (section \ref{subsec:mass}), 
$M_{\rm H\emissiontype{I}}=(2.9\pm0.2)\times10^9$ \MO  \citep{walter2008} and 
$M_{\rm H\emissiontype{II}}\leq1.4\times 10^9$ \MO (\cite{vanderhulst1988}, \cite{read2001}).
We obtained the inclination angle of $i = \timeform{22D} \pm \timeform{3D}$, using $V_{\rm rot} = 190 ^{+17}_{-12}$ km s$^{-1}$ derived from the baryonic mass (see  table~\ref{tab:mass}) and $V_{\rm gal,o}=70\pm7$~km~s$^{-1}$ in the range of $40'' \leq |Y|\leq 180''$ and $|X|\leq10''$ (see figure~\ref{fig:V_lsr}).
The derived inclination angle is consistent with the value of $i = \timeform{20D} \pm\timeform{5D}$ derived by a morphological analysis of H${\alpha}$ data \citep{tullyb1974}.
The radial variation of the inclination angle of the disk ($20'' \lesssim r \lesssim 105''$) was evaluated kinematically by \citet{shetty2007}, suggesting that the disk was warped and twisted. 
Actually, the rotation velocities are steeply decreased in the outer disk traced by HI (figure \ref{fig:V_lsr}).
This can indicate that the outer disk is warped and more inclined than the inner disk.

\begin{table}[tbp]
\caption{The baryonic masses of M51}
	\begin{center}
	\begin{tabular}{cccc}
	\hline
	\hline
	Baryon 											&mass (\MO)					& ratio(\%)			&reference\\\hline
	star												& 5.2 $^{+1.4}_{-1.1}\times 10^{10}$	&80.2				& HyperLeda database, \citet{bell2003}\\
	H\emissiontype{II}\footnotemark[$\ast$] (T$\sim$60000 K)	& $\leq9.5 \times 10^8$			& $\leq1.5$ 			& \citet{read2001}  \\
	H\emissiontype{II}\footnotemark[$\ast$] (T$\sim$5000K)		& 9.2 $\pm 0.9\times 10^8$		& 1.4 				& \citet{vanderhulst1988} \\
	H\emissiontype{I}\footnotemark[$\ast$]					& 3.9 $\pm 0.3\times 10^9$		&6.1 				&\citet{walter2008} \\
	H$_2$\footnotemark[$\ast$]							& 7.1 $\pm 0.9 \times 10^9$		& 10.9				& This paper \\\hline
	Total 											& $6.5 \times 10^{10}$\\
	\hline
	\hline
	$\ast$ including helium&&&\\
	\end{tabular}
	\end{center}
\label{tab:mass}
\footnotetext[$\ast$]{including helium}
\end{table}

\subsection{Mass and Radial Distribution of Molecular gas}
\label{subsec:mass}

\begin{figure}[tp]
\begin{center}
 \FigureFile(100mm,80mm){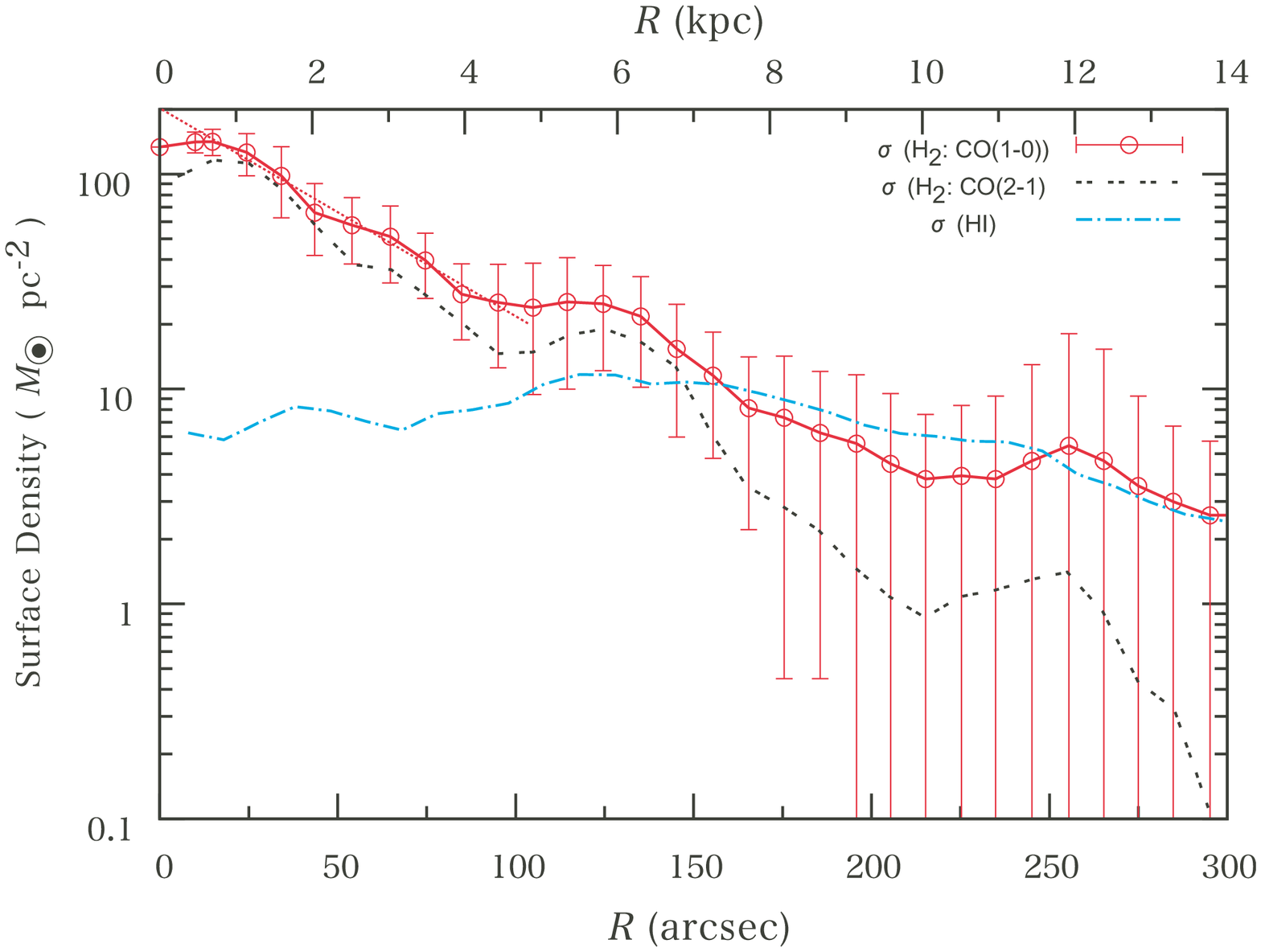}
 \end{center}
\caption{Radial distributions of the surface mass densities, $\sigma(\rm{H\emissiontype{I}})$ and  $\sigma\rm{(H_2)}$ derived from CO(1$-$0) and CO(2$-$1), including helium, 
where H\emissiontype{I} data of \citet{walter2008} and CO(2$-$1) data of \citet{schuster2007} were used, 
and $R=\sqrt{\mathstrut X'^2+Y^2}$ and $X'=X/\cos i$.
The surface density of the molecular gas decreases outward with an exponential shape fitted well by $\sigma [{\rm H_2, CO}(1$$-$$ 0)] = 202 \exp[-R / 2.0$~kpc]~\MO~pc$^{-2}$ at $1\leq R \leq4$~kpc (dotted line).
The increase at $R \approx $12~kpc shows the molecular gas in NGC 5195.}
\label{fig:rad_dis}
 \end{figure}

The H$_{2}$ surface density is obtained from $I_{\rm CO}$ by applying 
a CO(1$-$0)-to-$\rm H_{2}$ conversion factor of  
1 $\times 10^{20}$~cm$^{-2}$ (K~km~s$^{-1})^{-1}$ in M51 \citep{nakai1995}, 
\begin{eqnarray}
N({\rm H_{2}})  [{\rm cm^{-2}}]&=&1 \times 10^{20}~I_{\rm CO}~[{\rm K~km~s^{-1}}] ~\cos {\it i},
\label{eq:col}
\end{eqnarray}
which is equivalent to 
\begin{eqnarray}
 \sigma({\rm H_{2}})~[\MO~{\rm pc}^{-2}]&=& 1.59~I_{\rm CO} [{\rm K~km~s^{-1}}] ~\cos {\it i}.
\end{eqnarray}
Multiplied by 1.36 to include helium, mass of molecular gas in M51 and NGC 5195 were obtained 
to be M(H$_{2})=7.1  \times10^{9}$ \MO and $4.1 \times10^{8}$ \MO, respectively, from figure~\ref{fig:map}a.
Table~\ref{tab:mass} summaries the masses of stars and various gasses in M51.
The molecular gas mass is about twice as much as the neutral atomic gas mass, and total gas mass of ${\rm M(H\emissiontype{I})}+ {\rm M(H_2)}+{\rm M(H\emissiontype{II})}$ is about 16 \% of the star mass or 14 \% of the total baryonic mass.

Figure~\ref{fig:rad_dis} shows the radial distributions of 
$\sigma(\rm{H\emissiontype{I}})$ and  $\sigma\rm{(H_2)}$, including helium,  
which were derived by averaging into annuli with a width of  $10''$. 
In addition to $\sigma({\rm H}_2)$ derived from CO(1$-$0), 
we also show  $\sigma({\rm H}_2)$ evaluated from the data of CO(2$-$1) \citep{schuster2007} for reference, simply using same conversion factor of equation (\ref{eq:col}) which would be underestimation for CO(2$-$1), because the intensity of CO(2$-$1) is usually lower than that of CO(1$-$0) in normal spiral galaxies (e.g., \cite{nakai1994}).
The radial distribution of $\sigma (\rm H_2)$  from CO(1$-$0) shows an exponential decrease (dotted line) of 
$\sigma({\rm H_2}) = 202~\exp [-R / 2$~kpc]~\MO~pc$^{-2}$  at  $1\leq R \leq4$~kpc where figure~\ref{fig:rad_vel} shows the differential rotation.
At $R<1$~kpc where figure~\ref{fig:rad_vel} shows the rigid rotation, $\sigma\rm{(H_2)}$ dips below the value expected from the exponential shape.
Such radial distributions at $R \leq 4$~kpc can be seen in most non-barred  galaxies and be explained by inflow of molecular gas due to viscosity of the gas (Nishiyama et al. 2001).
At $4<R<10$~kpc, $\sigma\rm{(H_2)}$ shows a bump at $R\approx 6$~kpc and again decreases exponentially with the radius at $6<R<10$~kpc.
The local maximum at $R \approx 6$~kpc could be caused due to the kinks or fractures of the spiral arms (see section \ref{subsec:spiral}) by the interaction with the companion galaxy NGC 5195.
At $R \geq 6$~kpc, the spiral arms bend inward then outward (see figures~\ref{fig:map} and \ref{fig:pitch}), and thus the mean surface density at $R \approx$~6 kpc becomes larger.
The increase at $R \approx $ 12~kpc is due to the molecular gas in NGC 5195.
The above trend of the radial distribution of CO(1$-$0) is more conspicuous in CO(2$-$1), which traces warmer and denser molecular gas than CO(1$-$0).
The molecular gas is the dominant component [$\sigma ({\rm H_2}) > \sigma ({\rm H\emissiontype{I}})$] at $\sigma ({\rm H\emissiontype{I}})+\sigma ({\rm H_2}) \geq 20$~\MO~pc$^{-2}$, which has been seen in many galaxies (e.g., \cite{nishiyama2001}).

\subsection{Rotation Curve}
\label{sec:v_rot}

\begin{figure}[tp]
\begin{center}
 \FigureFile(100mm,80mm){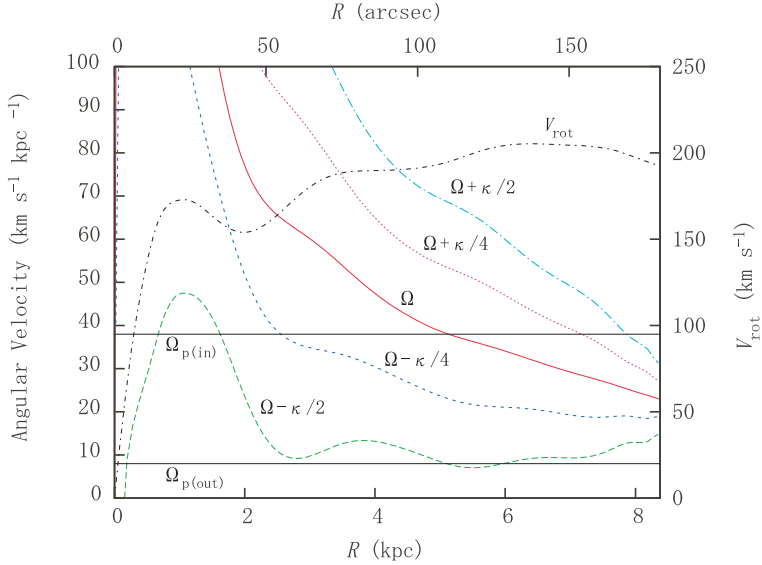}
 \end{center}
\caption{The rotation curve ($V_{\rm rot}$) and the angular velocities. 
The rotation curve is an average of the northern ($Y>0$) and southern ($Y<0$)  rotation velocities in figure~\ref{fig:V_lsr}, correcting the inclination of M51 ($\timeform{22D}$).
The angular velocities ($\Omega$, $\Omega\pm\kappa/2$, and $\Omega\pm\kappa/4$) are plotted as a function of the distance from the dynamical center of M51.  
The constant pattern speeds of $\Omega_{\rm p (in)}=38$ [km~s$^{-1}$~kpc$^{-1}$] and $\Omega_{\rm p (out)}=8$ [km~s$^{-1}$~kpc$^{-1}$] are also plotted. }
\label{fig:rad_vel}
 \end{figure}

Figure~\ref{fig:rad_vel} shows the rotation curve derived from the average of the northern ($Y>0$) and southern ($Y<0$) rotation velocities in figure~\ref{fig:V_lsr},
 using the inclination angle of $i=\timeform{22D}$ and the position angle of $\theta_{\rm PA}=\timeform{-10D}$.
The uncertainties of the inclination angle, $\Delta i=\pm\timeform{3D}$, and the position angle, $\Delta\theta_{\rm PA} \approx \pm\timeform{10D}$, influence the rotation velocity by $10$\% and $\sim2$\%, respectively.
The velocities in H$\alpha$ with the Kitt Peak 4-m telescope \citep{goad1979} 
and in CO with the BIMA \citep{shetty2007} were used in the region of $R \leq 30''$ and $R\leq40''$, respectively.
In the central region ($R < 15'' \approx$ 0.7~kpc), the rotation curve rises steeply like the rigid rotation. 
On the other hand, in the outer region, the velocity gradually increases from $R \approx 2.0$~kpc to $\approx 6.0$~kpc, showing the differential rotation with a dip at $R \approx 2.0$~kpc due to non-circular motion of the molecular gas influenced by the spiral potential (section \ref{subsec:pa}).

We obtained the angular velocities, $\Omega$, $\Omega\pm\kappa/2$ and $\Omega\pm\kappa/4$ from the rotation curve in order to estimate the locations of resonances in the orbits of stars and gas clouds, where $\kappa = \sqrt{\mathstrut R(d\Omega^{2}/dR)+4\Omega^{2}}$
 is an epicyclic frequency.
The locations of resonances are related closely to the structure of the spiral arms and the disk 
(e.g., \cite{lin1964}, \cite{binney1987}, 2008).
Adopted the single pattern speed of $\Omega_{\rm p} = 38$~km~s$^{-1}$~kpc$^{-1}$ (\cite{tullyc1974}, \cite{zimmer2004}), 
the radii of the inner (IILR) and the outer (OILR) inner Lindblad resonances, 
$\Omega_{\rm p}=\Omega-\kappa/2$ are about $R =0.7$~kpc ($15''$) and
1.6~kpc ($35''$), respectively.
We also find the co-rotation (CR; $\Omega=\Omega_{\rm p}$) at 5.1~kpc ($110''$), 
the outer Lindblad resonance (OLR; $\Omega_{\rm p}=\Omega+\kappa/2$) at 7.9~kpc ($169''$)
and the 4/1 resonance ($\Omega_{\rm p}=\Omega-\kappa/4$) at 2.6~kpc ($55''$). 
The OILR and CR nearly correspond to the positions of Tully ($R \rm \sim 1.9$~kpc, 1974b)   and Nikola ($R \rm \sim 5.6$~kpc, 2001), respectively.

Some previous studies (\cite{elmegreen1989}, \cite{salo2000b}, \cite{meidt2008}) have however suggested that the pattern speed of M51 varies with the radial distance due to the tidal perturbation induced by an encounter with the companion NGC 5195 (e.g.,\cite{toomre1972}).
\citet{elmegreen1989} proposed two pattern speeds, the inner spiral mode and the outer mode.
The former mode, adopting the pattern speed ($\Omega_{\rm p}=38$~km~s$^{-1}$~kpc$^{-1}$) of \citet{tullyc1974}, has an OLR at the position ($R \sim 8$~kpc) 
of a prominent intensity gap of the arms in M51 (figure~\ref{fig:map}a, figure~\ref{fig:pitch}). 
On the other hand, the latter mode is the material pattern that rotates with the companion, whose 
ILR coincides with CR of the inner mode. 
\citet{salo2000b} also demonstrated the variation of the pattern speed with the radius by a N-body simulation.
Using the outer pattern speed of $\Omega_{\rm p} \approx 8$~km~s$^{-1}$~kpc$^{-1}$ (e.g., \cite{salo2000b}), 
the ILR of the outer mode is coincident with the CR (5.1~kpc) of the inner mode ($\Omega_{\rm p} = 38$~km~s$^{-1}$~kpc$^{-1}$).
We hereafter adopt the pattern speed of 
$\Omega_{\rm p} = 38$~km~s$^{-1}$~kpc$^{-1}$ and 
8~km~s$^{-1}$~kpc$^{-1}$ 
in the range of $R = 40''$ - $110''$ and $110''$ - $140''$, respectively.

\subsection{Molecular Spiral Structure}

\begin{figure}[tp]
\begin{tabular}{ll}
 \begin{minipage}{0.5\hsize}
  \par
  {\textbf{(a)}}
 \begin{center}
  \FigureFile(80mm,100mm){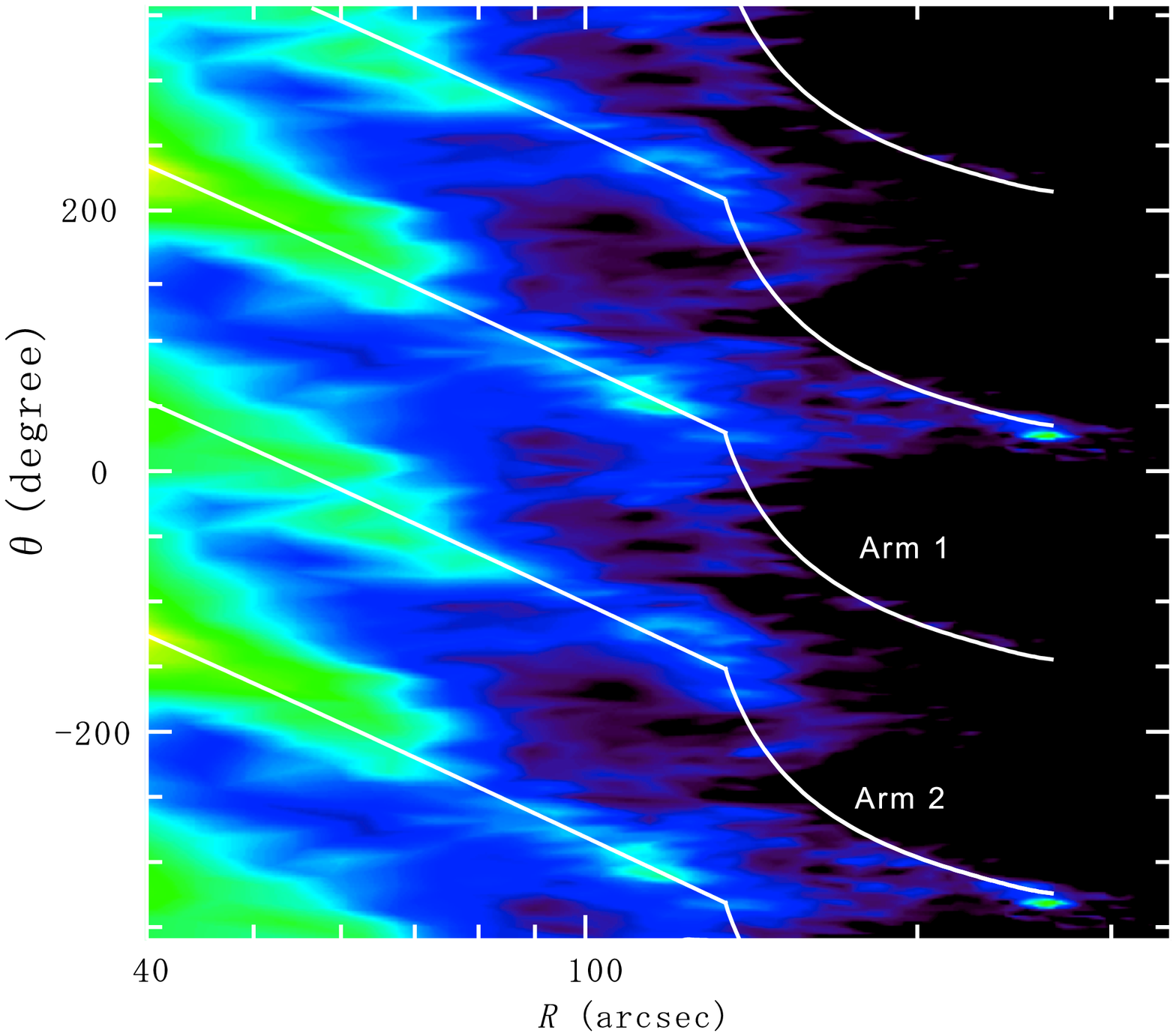}
   \end{center}
   \end{minipage}
 \begin{minipage}{0.5\hsize}
\par
{\textbf{(b)}}
 \begin{center}
 \FigureFile(60mm,80mm){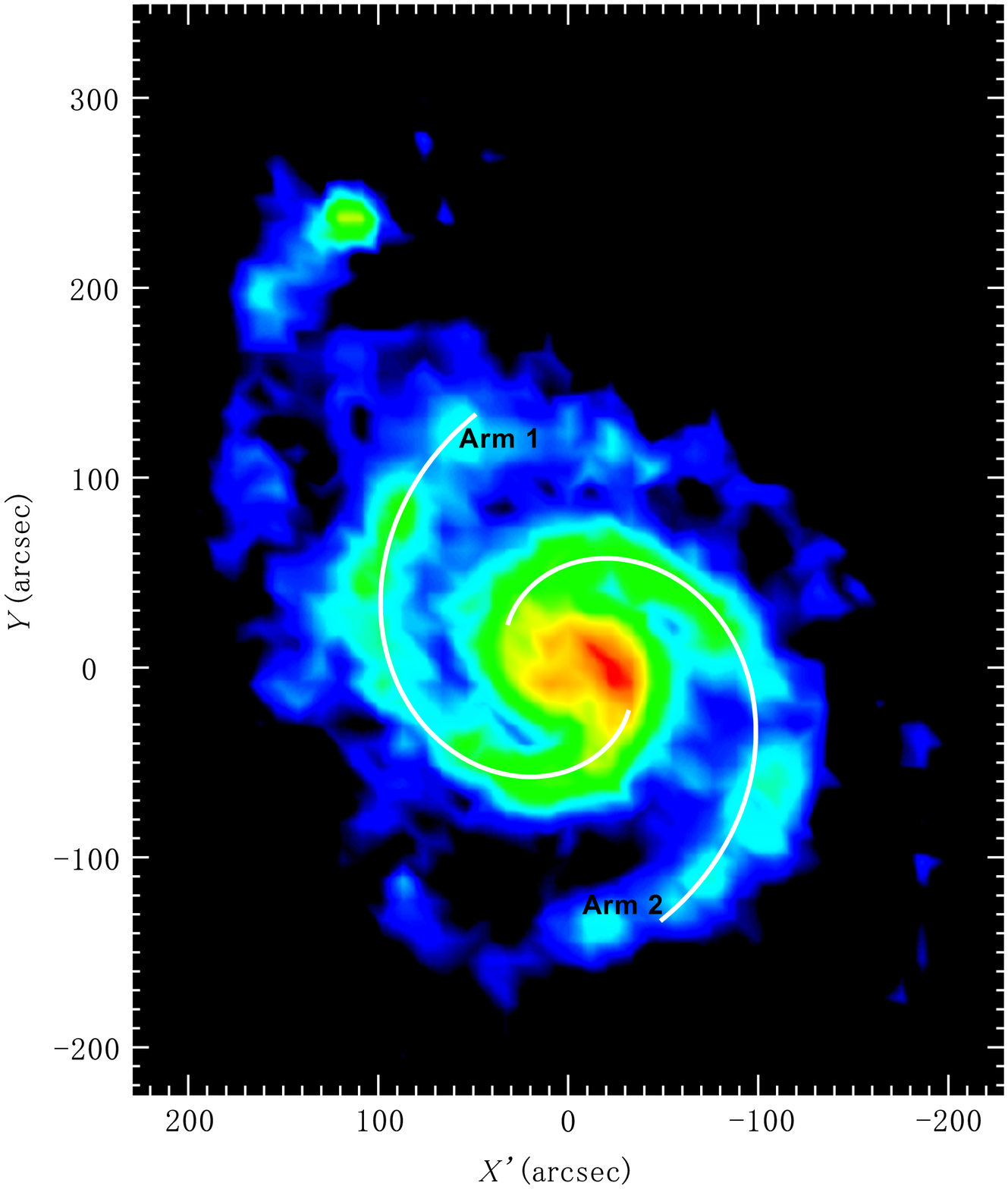}
 \end{center}
 \end{minipage}\\
 \end{tabular}
\caption{\textbf{(a)} ln $R - \theta$ plot of $I_{\rm CO}$ at $40'' \leq R \leq 340''$, where $\theta$ is measured counterclockwise from the major axis of the galactic disk ($\theta_{\rm PA} = -\timeform{10D}$) and the inclination angle (\textit{i} = $\timeform{22D}$) is corrected.  
Both arm 1 and 2 can be fitted by logarithmic spirals with a pitch angle of $\timeform{19D} \pm \timeform{1D}$ at $40'' \leq R \leq 140''$.\\
\textbf{(b)} Overlay of the logarithmic spirals with a pitch angle of $\timeform{19D}$ at $40'' \leq R \leq 140''$, separated by $\timeform{180D}$, on 
the $I_{\rm CO}$ map. 
The minor axis $X' (=X / \cos i)$
is corrected for the inclination angle.
}
 \label{fig:pitch}
\end{figure}

\label{subsec:spiral}
M51 has two very prominent spiral arms (figure~\ref{fig:map}) which are logarithmic
but are broken at some radius ($R\sim140''$), as pointed out by \citet{nakai1994}, 
and this feature also can be seen in the $\rm H{\alpha}$ image \citep{tullyc1974}.
This kinks can be caused by the tidal interaction with NGC 5195, 
which were reproduced by disk simulations (\cite{salo2000a}, \cite{dobbs2010}).
The angle between the tangent of a spiral arm and the azimuthal direction (i.e. pitch angle, \textit{p}) is represented by
\begin{equation}
-\tan p = \frac{1}{R} \frac{dR}{d\theta},
\label{eq:pitch}
\end{equation}
where 
$\theta$ is the azimuthal angle of the arm measured counterclockwise from the major axis of the galactic disk ($\theta_{\rm PA} = -\timeform{10D}$) 
and $R$ the radius corrected the inclination angle (\textit{i} = $\timeform{22D}$).
When \textit{p} is constant, equation (\ref{eq:pitch}) can be integrated as follows:
\begin{eqnarray}
\theta_0(R)
&=& \theta_{0} - \frac{1}{\tan p}~{\rm ln}\left(\frac{R}{R_{0}} \right), 
\label{eq:pitch_int}
\end{eqnarray}
where $R_{0}$ and $\theta_{0}$ are constants 
and $\theta_0(R)$ the position angle of a spiral arm at the radius $R$, measured from the major axis of the galactic disk.
Figure~\ref{fig:pitch}a shows a ln$R - \theta$ plot of the $I_{\rm CO}$ in figure~\ref{fig:map}a corrected the inclination angle. 
The spiral arms can be fitted by equation (\ref{eq:pitch_int}) at $40'' \leq R \leq 140''$, with a pitch angle of $p = \timeform{19D} \pm\timeform{1D}$, 
where $R_{0} = 40''$ and $\theta_{0} = \timeform{230D}$. 
The pitch angle agrees with the previous values $p=\timeform{15D}\pm\timeform{4D}$ \citep{elmegreen1989} and $p=\timeform{21D}\pm\timeform{5D}$ \citep{nakai1994}.
At $R > 140''$, the two spiral arms are broken.



\section{Discussion}
\label{chap:res}
\subsection{Derivation of Velocity Vectors}
\label{subsec:velvec}

\begin{figure}[tp]
\begin{tabular}{lr}
 \begin{minipage}{0.5\hsize}
  \par
  {\textbf{(a)}}
 \begin{flushleft}
 \FigureFile(80mm,50mm) {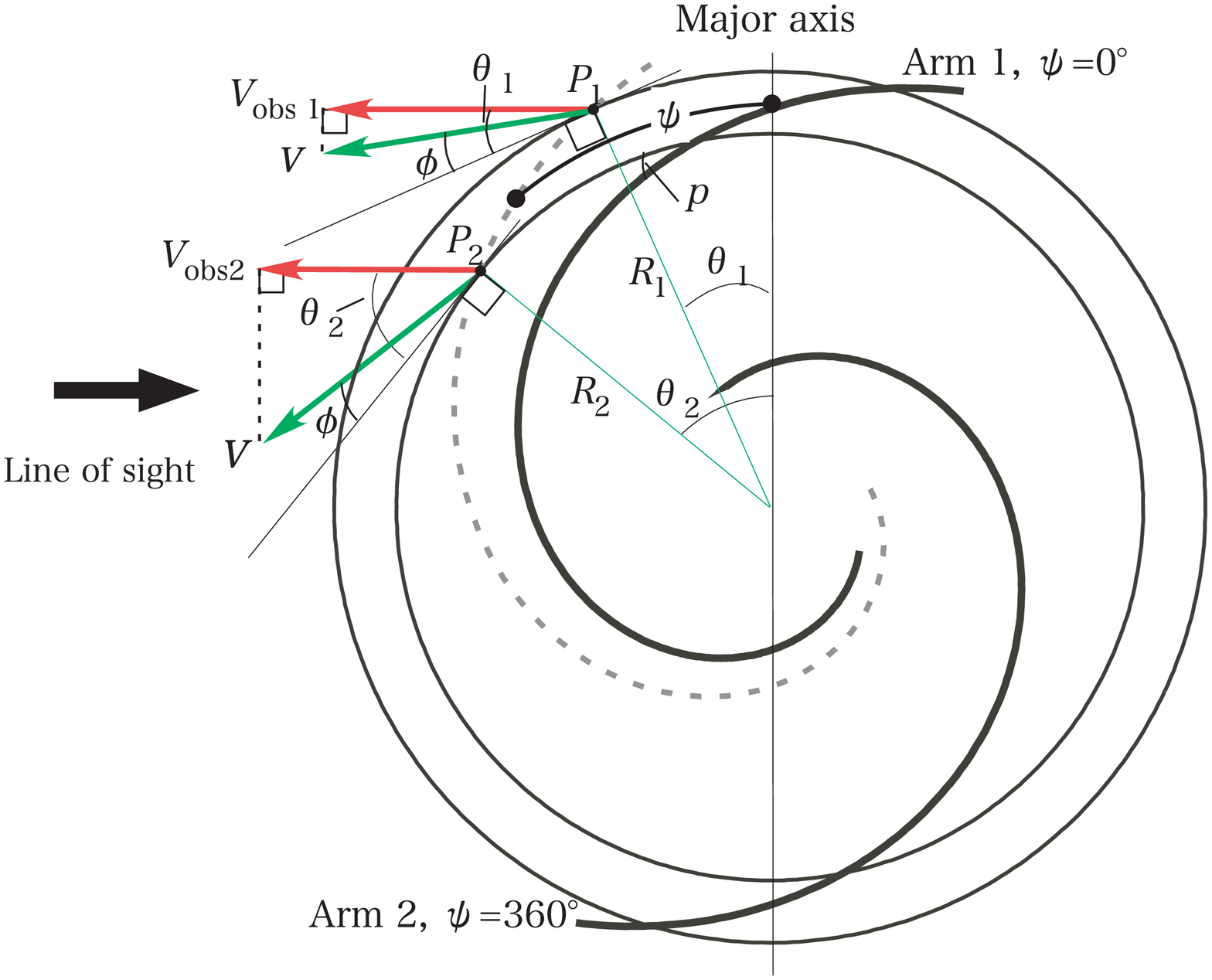}
   \end{flushleft}
   \end{minipage}
 \begin{minipage}{0.5\hsize}
\par
{\textbf{(b)}}
 \begin{flushright}
 \FigureFile(80mm,50mm) {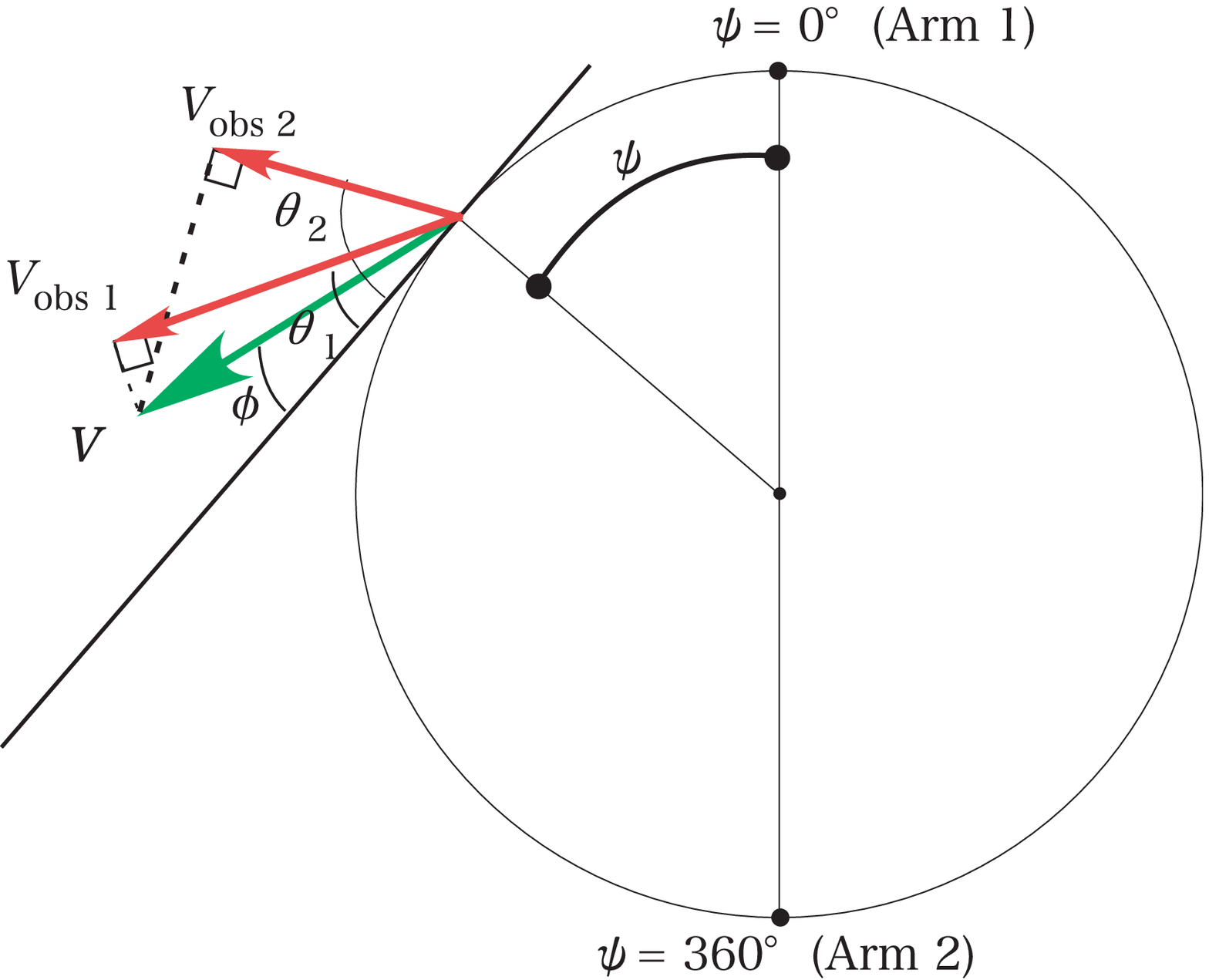}
 \end{flushright}
 \end{minipage}\\
 \end{tabular}
\caption{The pattern diagram of the spiral phase. 
\textbf{(a)} The phase of arm 1 starts from $\Psi = \timeform{0D}$, and arm 2 from $\Psi = \timeform{360D}$. 
The velocity vectors $\vec{V} = (V,$ $\phi)$ at the same spiral phase are regarded as same in the region between the radii $R_{1}$ and $R_{2}$ which are not great different each other. 
We see the same velocity vector $\vec{V} = (V, \phi)$ located at the spiral phase $\Psi$ 
in two different directions at $P_1=(R_1$, $\theta_1$) and $P_2=(R_2$, $\theta_2$) and hence can measure 
two different components of the velocity vector, 
$V_{\rm obs 1} = V \cos(\theta_1 - \phi)$ and $V_{\rm obs 2} = V \cos(\theta_2 - \phi)$.
\textbf{(b)} The velocity vector $\vec{V} = (V, \phi)$ at the spiral phase $\Psi$ can be determined from 
the two different velocity components $V_{\rm obs 1}$ and $V_{\rm obs 2}$ measured in the line of sight.
If we measure more than two components of the velocity vector, the velocity vector can be determined 
statistically more accurately.}
 \label{fig:ponchi}
\end{figure}

Figure~\ref{fig:map}b shows a contour map of the CO velocity field with the gas approaching us in the northern part of the disk of M51 and receding in the south.
The distorted isovelocity contours indicate the disturbed motion from pure circular rotation.
Large non-circular motions in the spiral arms, regarded as the streaming motion of interstellar gas, have been detected in $\rm H{\alpha}$ \citep{tullyc1974} and CO (\cite{vogel1988};  \cite{garcia1993a}, b; \cite{rand1993}; \cite{kuno1997}; \cite{aalto1999}).
It is difficult however to know variations of velocity vectors in the galactic disk, 
because we can measure only the velocity component in the line of sight.
In order to investigate the motion and orbit of interstellar gas in the spiral potential,
\citet{kuno1997} proposed a method (KN method) to derive the velocity vectors in the spiral phases from measured velocity data.
Figure~\ref{fig:ponchi} shows its technique to obtain a velocity vector. 
It is regarded that 
the velocity vectors, $\vec{V}$, located at the same spiral phase ($\Psi$; dotted line in figure~\ref{fig:ponchi}a) in the nearly same radius (e.g., $R_{a} < R < R_{b}$ ) are same.
Under the assumption, we can observe different components of a velocity vector at a spiral phase in the line of sight (figure~\ref{fig:ponchi}a), and thus the velocity vector can be determined using at least two different components observed (figure~\ref{fig:ponchi}b).

The observed line-of-sight velocities $V_{\rm obs 1}$ and $V_{\rm obs 2}$ at  two different observed positions $P_1$ and $P_2$ at $R_1$ and $R_2$ ($R_a < R_1$, $R_2<R_b$) in figure \ref{fig:ponchi} 
are given by 
\begin{equation}
V_{\rm obs 1} = V_{\rm sys}-V \cos(\theta_1-\phi) \sin i, 
\label{eq:velvec1}
\end{equation}
\begin{equation}
V_{\rm obs 2} = V_{\rm sys}-V \cos(\theta_2-\phi) \sin i,
\label{eq:velvec2}
\end{equation}
 where $\theta_1$ and $\theta_2$ are the position angle of the positions measured from the major axis in the galactic plane,
\textit{i} the inclination angle of the galactic plane, and $V_{\rm sys}$ the systemic velocity of the galaxy.
$V$ and $\phi$ represent the amplitude of the velocity vector and the offset angle from the tangential direction of a circular orbit, respectively (figure \ref{fig:ponchi}a).
Two measured line-of-sight velocities $V_{\rm obs 1}$ and $V_{\rm obs 2}$ determine two unknown parameters
$V$ and $\phi$ by solving equations (\ref{eq:velvec1}) and (\ref{eq:velvec2}), where $V_{\rm sys}$ and $i$ are given in section \ref{subsec:para}.
Thus we can know the velocity vector $\vec{V}=(V$, $\phi$) at the spiral phase $\Psi$.
In other words, as shown in figure \ref{fig:ponchi}b, the velocity vector $\vec{V}=(V$, $\phi)$ can be determined from two different components $V_{\rm obs 1}$ and $V_{\rm obs 2}$ of the velocity vector at the spiral phase $\Psi$.
If there are more than two line-of-sight velocities observed at the same phase $\Psi$ at $R$ ($R_a<R<R_b$), 
the velocity vector $\vec{V}=(V$, $\phi)$ can be determined more accurately.
If the velocity vector $\vec{V}=(V$, $\phi$) at $R_i$ ($R_a<R_i<R_b$) however is not same even at the same spiral phase $\Psi$, the evaluated velocity $\vec{V}$ has errors, or indicates the mean of the velocity, $\langle {V}  \rangle$, and the offset angle, $\langle \phi  \rangle$, at $R_a<R_i<R_b$, whose dispersions are large.

\begin{figure}[tp]
\begin{tabular}{cc}
 \begin{minipage}{0.5\hsize}
  \par
    {\textbf{(a)}}
 \begin{flushleft}
 \FigureFile(80mm,50mm) {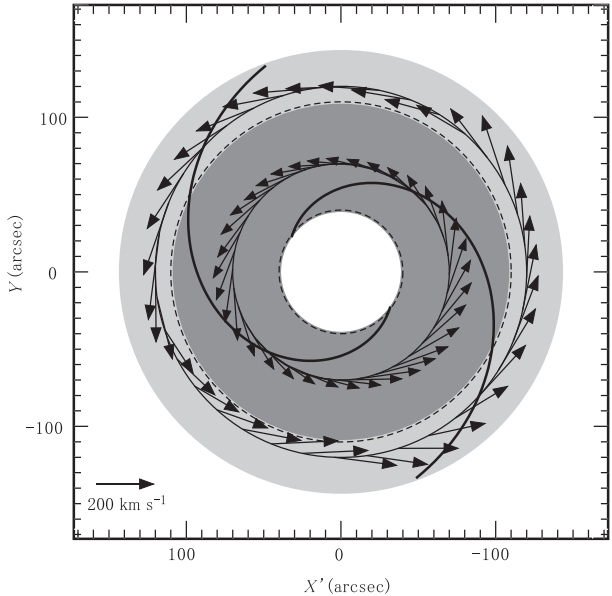}
   \end{flushleft}
   \end{minipage}
 \begin{minipage}{0.5\hsize}
\par
  {\textbf{(b)}}
 \begin{flushright}
 \FigureFile(65mm,40mm) {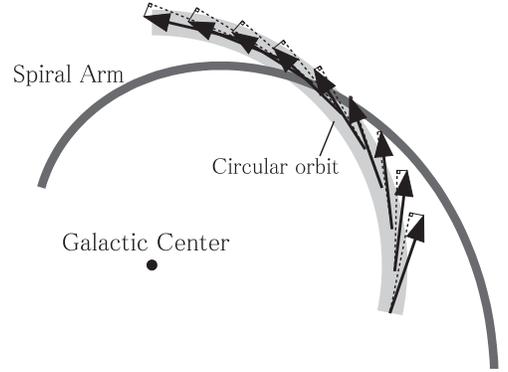}
 \end{flushright}
 \end{minipage}\\
 \end{tabular}
\caption{\textbf{(a)} The representative velocity vectors in $R = 40''$ - $110''$ (dark gray) 
and $R = 110''$ - $140''$ (light gray) and the spiral phases $\Psi$ with $\Delta \Psi = \timeform{20D}$.
\textbf{(b)}  Enlargement of the velocity vectors across the arm, $\Psi = \timeform{250D}$ - \timeform{410D}, in $R=40''$ - $110''$ around ($X'$, $Y$)$\sim(-50''$, $60''$). 
The dotted lines show the tangential direction of a circular orbit (light gray line).
The velocity vectors change their direction from outward to inward in the arm.
}
\label{fig:stream}
\end{figure}

\begin{figure}[p]
\begin{center}
\begin{tabular}{cc}
 \begin{minipage}{0.45\hsize}
  \par
 \begin{center}
$R= 40''$ - $110''$
  \FigureFile(70mm,150mm){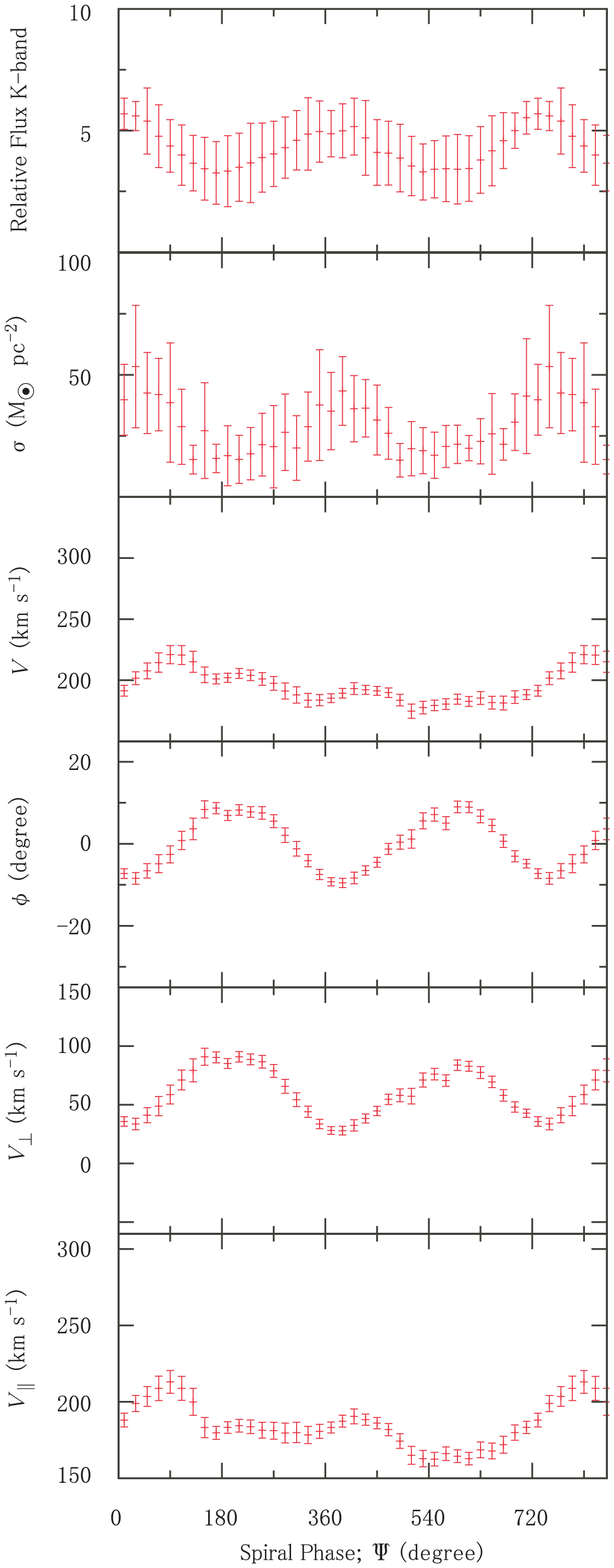}
   \end{center}
   \end{minipage}
 \begin{minipage}{0.45\hsize}
\par
\begin{center}
$R = 110''$ - $140''$
\FigureFile(70mm,150mm){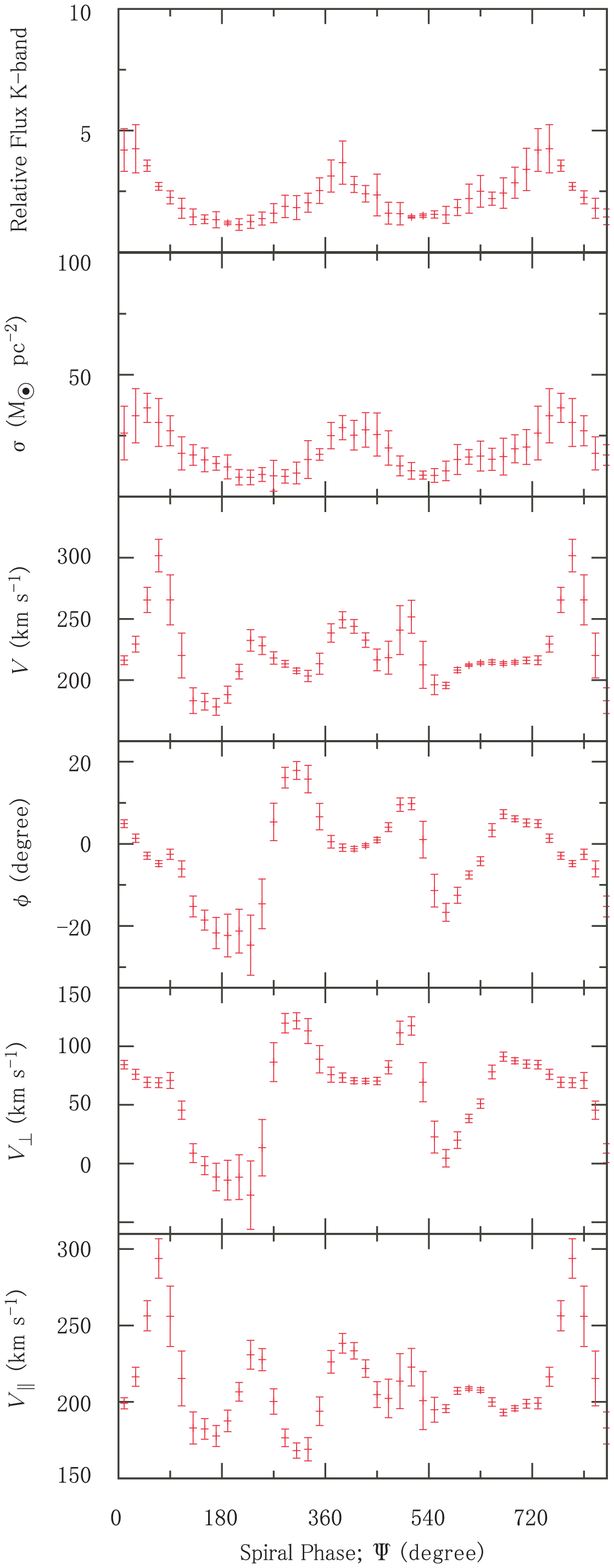}
 \end{center}
 \end{minipage}
 \end{tabular}
\caption{The relative flux density of K-band, the surface density of molecular gas (H$_2$), the amplitude of the velocity vector, the offset angle from the tangential direction of a circular orbit 
and the parallel ($V_{\parallel}$) and the perpendicular ($V_{\perp}$) velocity to the arms in the spiral phases $\Psi$, from the top panel to the bottom panel.
The left and right panels represent the regions in $R = 40''$ - $110''$ and $110''$ - $140''$, respectively. 
The spiral phases $\Psi$ in the regions are measured counterclockwise from $\theta_{\rm PA} = -\timeform{5D}$ and $ -\timeform{15D}$, respectively.}
 \label{fig:velvec}
\end{center}
\end{figure}

We applied this method to our new data of figure~\ref{fig:map}b. 
Although \citet{kuno1997} regarded the velocity vectors in the wide range of the radius ($R_a=\rm40'' <R < R_b=140''$) as constant values,
we derived the velocity vectors at the radii subdivided into $R= 40''$ - $110''$ and $110''$ - $140''$, 
because the depth of the spiral potential would vary with radius and hence the motion and the orbit of gas would also vary.
In the range of $R = 40''$ - $140''$, 
the spiral arms can be fitted by the logarithmic spiral with the same pitch angle, $p = \timeform{19D}$ (section \ref{subsec:spiral}), 
but the adopted pattern speed is different at $R < 110''$ and $R > 110''$ (section~\ref{sec:v_rot}).
Next, these regions were divided into the spiral phase width of $\Delta \Psi = \timeform{20D}$ 
(actual angular width of $\Delta\Psi/2=\timeform{10D}$ whose spacing of $\sim 650$~pc at $R=80''$ matches our observational angular resolution).
The spiral phase $\Psi (R, \theta)$ is defined by 
\begin{eqnarray}
\Psi(R, \theta)  & = &  2\times \left[\theta(R)-\theta_{0}(R)\right],
\label{eq:spiral_phase}
\end{eqnarray}
where $\theta(R)$ is the position angle of the observed point at the radius $R$, measured from the major axis of the galactic disk (figure~\ref{fig:ponchi}a). 
We derived $V$ and $\phi$ by a least-squares fit using the observed velocities at the various position angles in a same spiral phase, using the inclination angle of $i = \timeform{22D}$ (section \ref{subsec:incl}) and the position angle of $\theta_{\rm PA} = -\timeform{5D}$ $(40'' \leq R \leq 110'')$ and $-\timeform{15D}$ $(110'' \leq R \leq 140'')$ (section \ref{subsec:pa}). 
The uncertainty of the inclination angle, $\Delta i=\pm\timeform{3D}$,
gives the uncertainty of $\Delta V/V\sim10$\%.
Figure~\ref{fig:stream}a shows the resultant velocity vectors in each spiral phase in the range of $40'' \leq R \leq 110''$ and $110'' \leq R \leq 140''$.
The velocity vectors in the range of $40''\leq R\leq110''$ change their direction with respect to a circular orbit from inward to outward in the upstream region of a spiral arm and from outward to inward in the arm (see figure \ref{fig:stream}b and $\phi$ in figure \ref{fig:velvec}) as expected by the density wave theory (\cite{levinson1981}, \cite{roberts1987}, \cite{roberts1990}). 
At the outer region of $X' \approx 0''$ - $100''$ and $Y\approx -50''$ - $-120''$, however, the direction is extremely disturbed (see section \ref{subsec:spiral}).

In order to investigate the velocity variations with respect to the spiral potential, 
we divided the velocity vectors into two components parallel ($V_{\parallel}$) and perpendicular ($V_{\perp}$) to the arms, 
$V_{\parallel}=V \cos(\phi + \textit{p})$ and $V_{\perp}=V \sin(\phi + \textit{p})$ where $p$ is the pitch angle of $\timeform{19D}$ (section \ref{subsec:spiral}) (see figure~\ref{fig:ponchi}a).
Figure~\ref{fig:velvec}  shows the variations of the velocity components seen in each corotating frame of the spiral arms, where we adopted the pattern speeds of $\Omega_{\rm p}=38$~km~s$^{-1}$~kpc$^{-1}$ and 8~km ~s$^{-1}$~kpc$^{-1}$ at $R = 40'' $ - $110''$ and $110''$ - $140''$, respectively (section~\ref{sec:v_rot}). 
The azimuthal variations of the surface density of the molecular gas and the relative flux in the K-band \citep{jarrett2003} which indicates the surface density of stars are also displayed for comparison.
In the inner region of $R=40''$ - $110''$, 
the variations of these velocity components with respect to the surface density are consistent with the results of \citet{kuno1997}, 
and are in agreement with the particle simulations (\cite{levinson1981}, \cite{roberts1987}, \cite{roberts1990})
which insist on the decline of the perpendicular velocity components $V_{\perp}$ 
across the arms ($-2 \lesssim {\Delta V_{\perp}}/{\Delta \Psi} \lesssim -1~[{\rm km}~{\rm s^{-1}}$ degree$^{-1}$])
caused by a type of shock due to collisions between gas clouds entering the arms, 
although figure~\ref{fig:velvec} is not as sharp as expected in the simulations for the gas as fluid (${\Delta V_{\perp}}/{\Delta \Psi} \ll -10~[{\rm km}~{\rm s^{-1}}$ degree$^{-1}$], see figure 5 of \citet{roberts1969}).

In the outer regions of $R= 110''$ - $140''$, 
however, the variations of these velocity components are different from those expected in the simulations.
The inclination of the outer warped disk can influence the variations of the derived velocity components as shown in rotation curve (section \ref{sec:v_rot}).
The motions perpendicular to the arms ($V_{\perp}$) are rapidly accelerated in the downstream of the arms ($\Psi =  \timeform{490D}$ - $\timeform{510D}$), while the accelerated motions in the upstream of the arms decrease in the arms as well as the inner regions.
Also $V_{\perp}$ in the interarms ($\Psi =\timeform{130D}$ - $\timeform{150D}$) shows suddenly deceleration.
The variations are  particular striking in the amplitude of  the velocity vectors ($V$).
The positions of $\Psi = \timeform{130D}$ - $\timeform{150D}$ and $\Psi = \timeform{490D}$ - $\timeform{510D}$ in the outermost regions  correspond to east side and west side of the M51 disk (e.g., $(X', Y') \approx (-115'', -30'')$ and $(115'', 30'')$ at $R = 110''$ - $140''$), respectively.
According to  Salo's simulation model (2000a) which accounts for not only the morphological feature of the M51 system but also observational kinematics, 
the azimuth of the companion NGC 5195 across the M51 disk is $\theta_{\rm PA}=-\timeform{15D}$, where the azimuth is counted counterclockwise from the major axis of $\theta_{\rm PA}=\timeform{170D}$.
They suggest that the materials in the outer regions are perturbed by multiple-encounter with NGC 5195 and have out-of-plane velocities.
As suggested by \citet{salo2000a}, in the case that the outer disk is tilted at $\sim \timeform{50D}$ with the opposite to the inner disk ($i=\timeform{22D}$), 
the effect of the inclination on the projected velocity is about $30 \%$.
The increasing and decreasing of $V$ at $\Psi = \timeform{130D}$ - $\timeform{150D}$ and $\Psi = \timeform{490D}$ - $\timeform{510D}$ can be affected by the out-of-plane velocities. 
This suggests that the molecular gas in the outermost region is no longer able to move as expected from the density wave theory. 
This view is consistent with the contention of \citet{elmegreen1989} and \citet{tullyc1974} who suggested the existence of the tidal spiral arms in the outer disk as a consequence of the tidal interaction with NGC 5195.

\subsection{Orbits of the Molecular Gas}
\label{subsec:orbit}

\begin{figure}[tp]
\begin{center}
 \FigureFile(80mm,50mm) {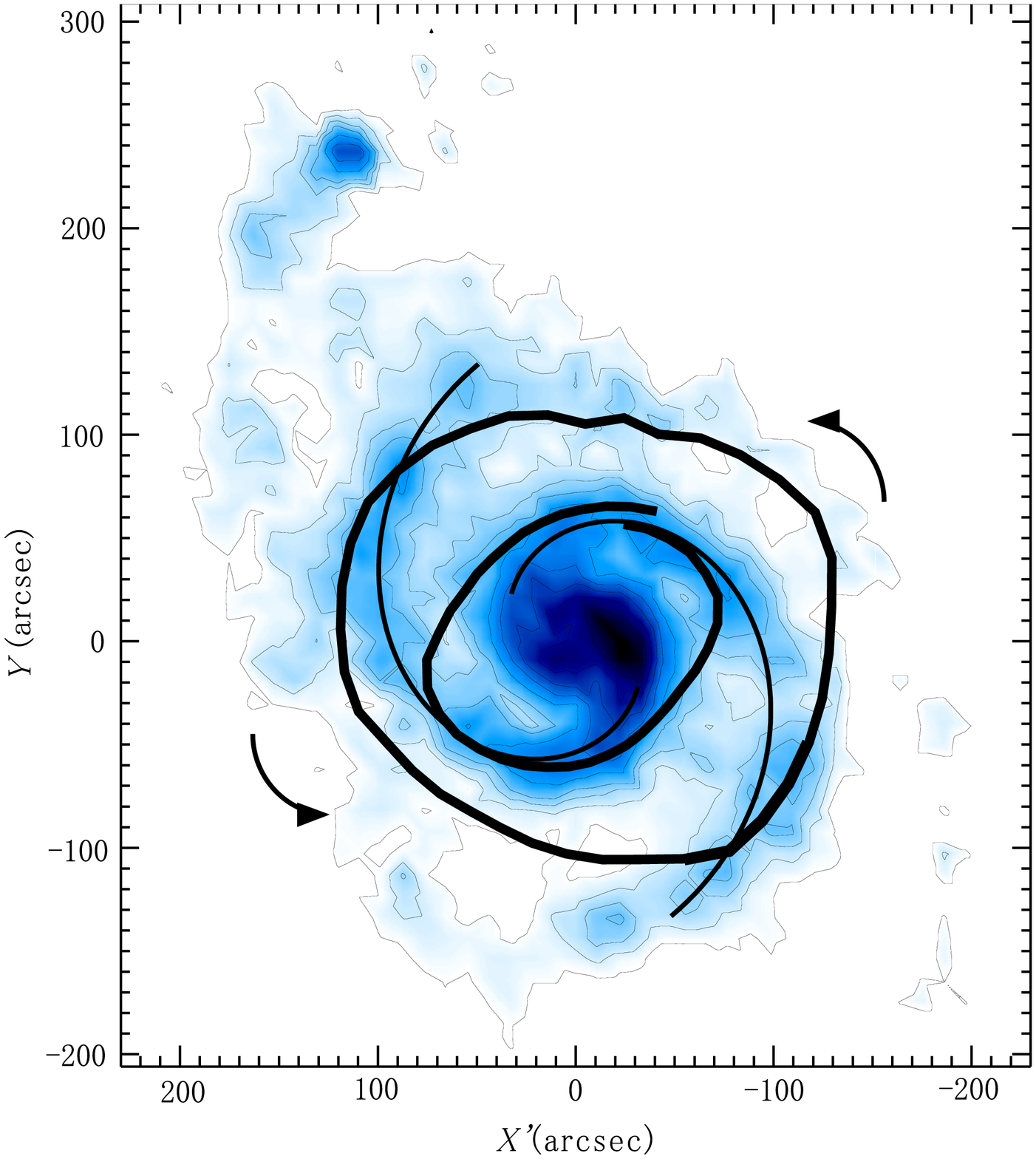}
\caption{Stremlines of the molecular gas derived from the velocity vectors at $R = 40''$ - $110''$ and $110''$ - $140''$ in figure~\ref{fig:stream}a overlaid on the CO integrated intensity of figure~\ref{fig:map}.
The frame corotates with the spiral pattern in each region. 
The direction of the rotation of the gas is counterclockwise.}
\label{fig:orbit}
\end{center}
\end{figure}

The streamline of the molecular gas in the frame corotating with the spiral pattern is  derived from the velocity vectors in figure~\ref{fig:stream}a, 
using the equations (6), (7) and (8) in  \citet{kuno1997}.
Figure~\ref{fig:orbit} shows the resultant orbits of the molecular gas, 
where the pattern speeds  $\Omega_{\rm p}$ are 38~km~s$^{-1}$~kpc$^{-1}$ and 8~km~s$^{-1}$~kpc$^{-1}$ at $R = 40'' $ - $110''$ and $110''$ - $140''$, respectively  (section~\ref{sec:v_rot}).
At $R = 40'' - 110''$, 
the radial velocities ($V_{\rm r}=V\sin\phi$) decelerate when the gas goes from the upstream side of the arms into the arms (figure \ref{fig:velvec}), 
and therefore the orbit changes along the spiral arms after the gas enters into the arms. 
As a result, the orbit of the molecular gas becomes an oval as 
predicted by the density wave theory.
 
In the outer region of $R=110''$ - $140''$, however, the molecular gas does not move along the arms, 
because the radial velocities randomly have negative values even in the interarms (figure \ref{fig:velvec}).
Also, the radial velocities reach minima in the interarms, 
whereas those in the inner region of $R=40''-110''$ minima in the arms. 
The orbit deviates from the shape expected from the density wave theory.
The deviation is associated with 
the motion of the gas in the outer disk as in the material arms rather than in the density wave, 
caused by the tidal interaction with NGC 5195 (e.g., \cite{tullyc1974}, \cite{elmegreen1989}).

\subsection{Destruction of Giant Molecular Associations by Shear Effect}
\label{subsec:shear}

\begin{figure}[tp]
\begin{center}
 \FigureFile(80mm,50mm) {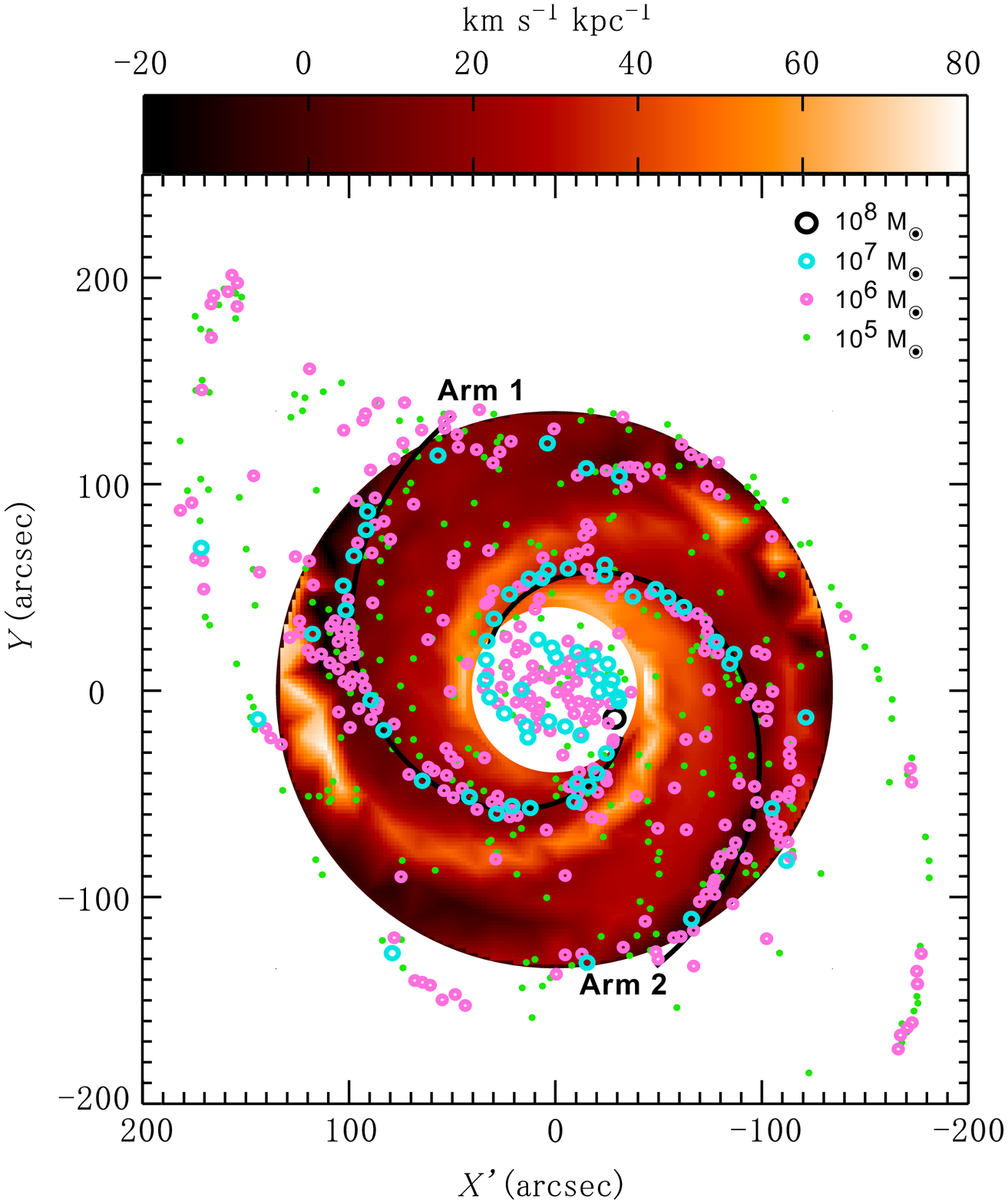}
\caption{The distribution of the shear strength (colour) derived from the velocity vectors in figure~\ref{fig:stream}a, overlaid on the distribution of GMAs  ($\geq 10^7$~\MO) and GMCs  ($\leq 10^6$~\MO) (open circles) shown by Koda et al. (2009). $X'$ and $Y$ are the minor axis corrected for the inclination angle and the major axis, respectively.}
\label{fig:shear}
\end{center}
\end{figure}
\begin{figure}[tp]
\begin{center}
\begin{tabular}{cc}
 \begin{minipage}{0.5\hsize}
  \par
 \begin{flushleft}
 \FigureFile(80mm,50mm) {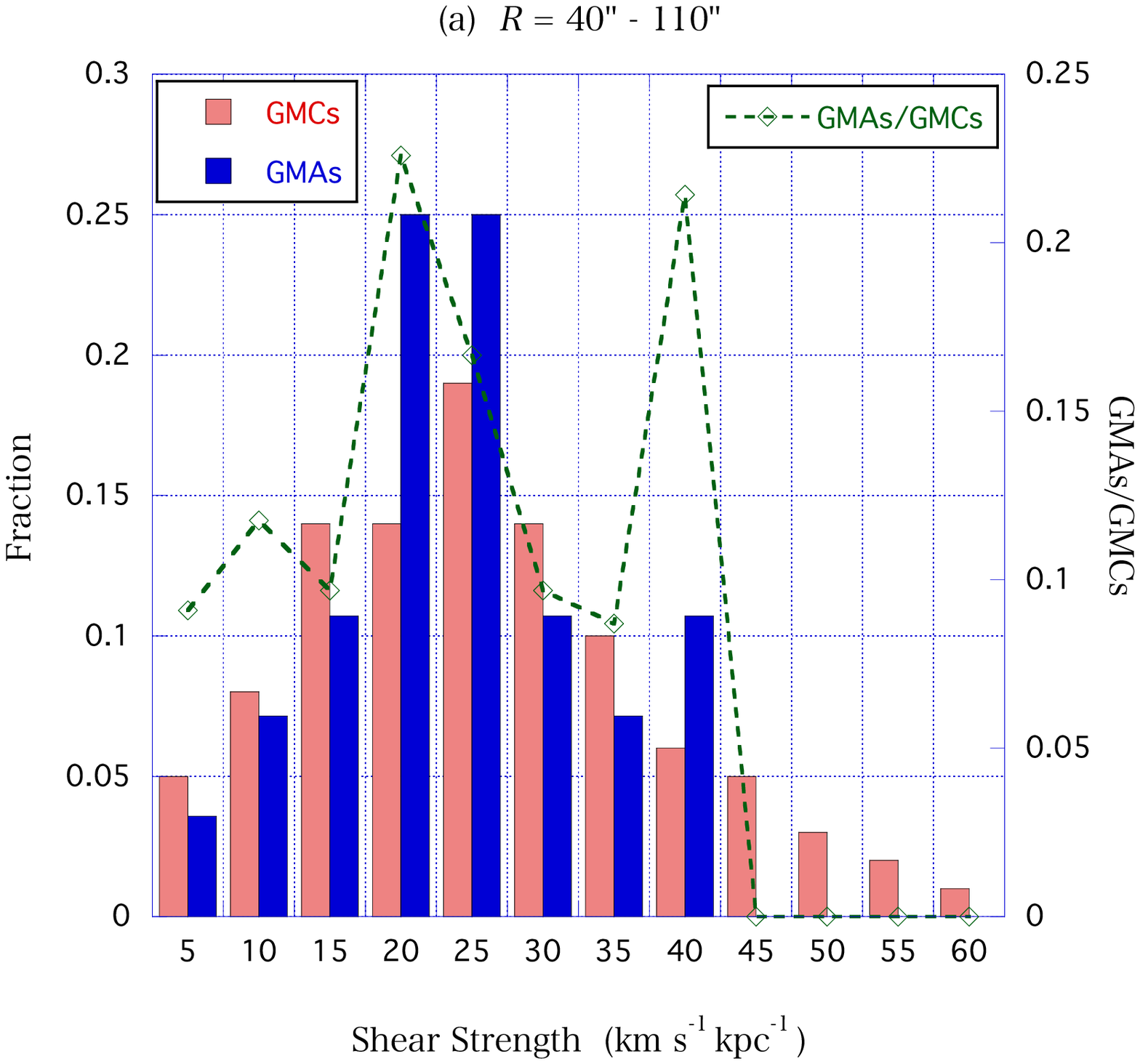}
   \end{flushleft}
   \end{minipage}
 \begin{minipage}{0.5\hsize}
\par
 \begin{flushright}
 \FigureFile(80mm,50mm) {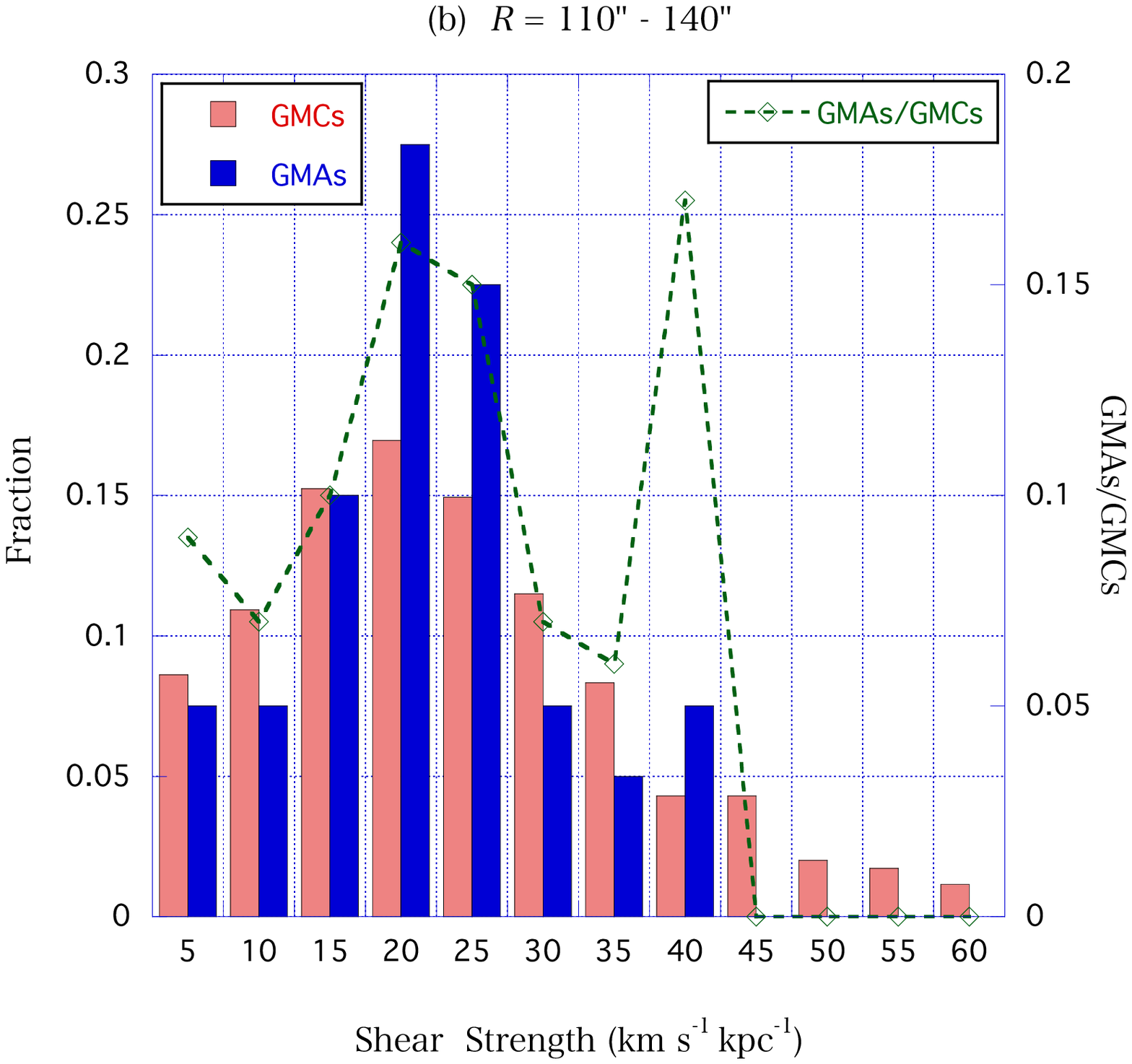}
 \end{flushright}
 \end{minipage}
 \end{tabular}
\caption{The ratios of the number of GMC (M $\sim 10^5$ - $10^6$ \MO; red bars) and of GMA (M $\sim 10^7$ - $10^8~\MO$  blue bars) in each shear strength to the total number of GMC ($f_{\rm GMC}=N_{\rm GMC}/\Sigma N_{\rm GMC}$) and of GMA ($f_{\rm GMA}=N_{\rm GMA}/\Sigma N_{\rm GMA}$), respectively, in $R=40''$-110$''$ (a) and $R=110''$-140$''$ (b). 
The line chart is the ratio of GMA to GMC in each shear strength.
}
 \label{fig:shear2}
\end{center}
\end{figure}

The variation of the streaming motion of the molecular gas leads to the radial gradient of the tangential velocity, i.e. shear.
The Oort's $A$-constant,  
\begin{eqnarray}
A  & = &  \frac{1}{2} \left(\frac{V_{\rm t}}{R} - \frac{dV_{\rm t}}{dR} \right)\nonumber \\ 
 & = &  - \frac{1}{2} R \frac{d \Omega}{dR}.
\end{eqnarray}
represents the rate of the shear in the disk (\cite{elmegreen1988}) and can be derived from velocity vectors ($V_{\rm t} = V\cos\phi$).
Figure~\ref{fig:shear} shows the distribution of the shear strength, $A$, derived from the velocity vectors in figure~\ref{fig:stream}a with the distribution 
of giant molecular clouds (GMCs, M $\sim$ $10^5$ - $10^6$ \MO) and giant molecular associations (GMAs, M $\sim$ $10^7$ - $10^8$ \MO)
which were identified by \citet{koda2009}.
The figure shows the anticorrelation between the distribution of the shear strength and of GMAs.
Namely GMAs exist only in the area of the weak shear strength and further on the upstream side 
of high shear strength, while GMCs exist on the downstream side 
and in the interarms.
Histograms in figure~\ref{fig:shear2} show the ratio of the number of GMCs or GMAs in each range of the shear strength 
from $A=0$ km s$^{-1}$ kpc$^{-1}$ to $60$ km s$^{-1}$ kpc$^{-1}$, 
whose interval is $\Delta A = 5$ km s$^{-1}$ kpc$^{-1}$, 
to the total number of GMCs or GMAs.
The red and  blue bars show the fractions of GMCs and GMAs, respectively.
There is a cutoff of the number counts of GMAs at the shear of $ 40$ km s$^{-1}$ kpc$^{-1}$ 
in contrast to the extended distribution of GMCs at $> 40$~km~s$^{-1}$~kpc$^{-1}$.  
The positions of GMCs and GMAs (figure~\ref{fig:shear}) and the fractions of GMCs and GMAs with respect to the shear strength (figure~\ref{fig:shear2}) strongly suggest that GMAs are fragmented at the downstream side of the arms by the strong shear,  
and then are released into the interarms as smaller clouds of GMC.

\subsection{Probability for Formation of Giant Molecular Associations}
\label{subsec:timescale}
\begin{figure}[tp]
\begin{center}
\begin{tabular}{cc}
 \begin{minipage}{0.5\hsize}
  \par
 \begin{flushleft}
  {\textbf{(a)}}
 \FigureFile(80mm,50mm) {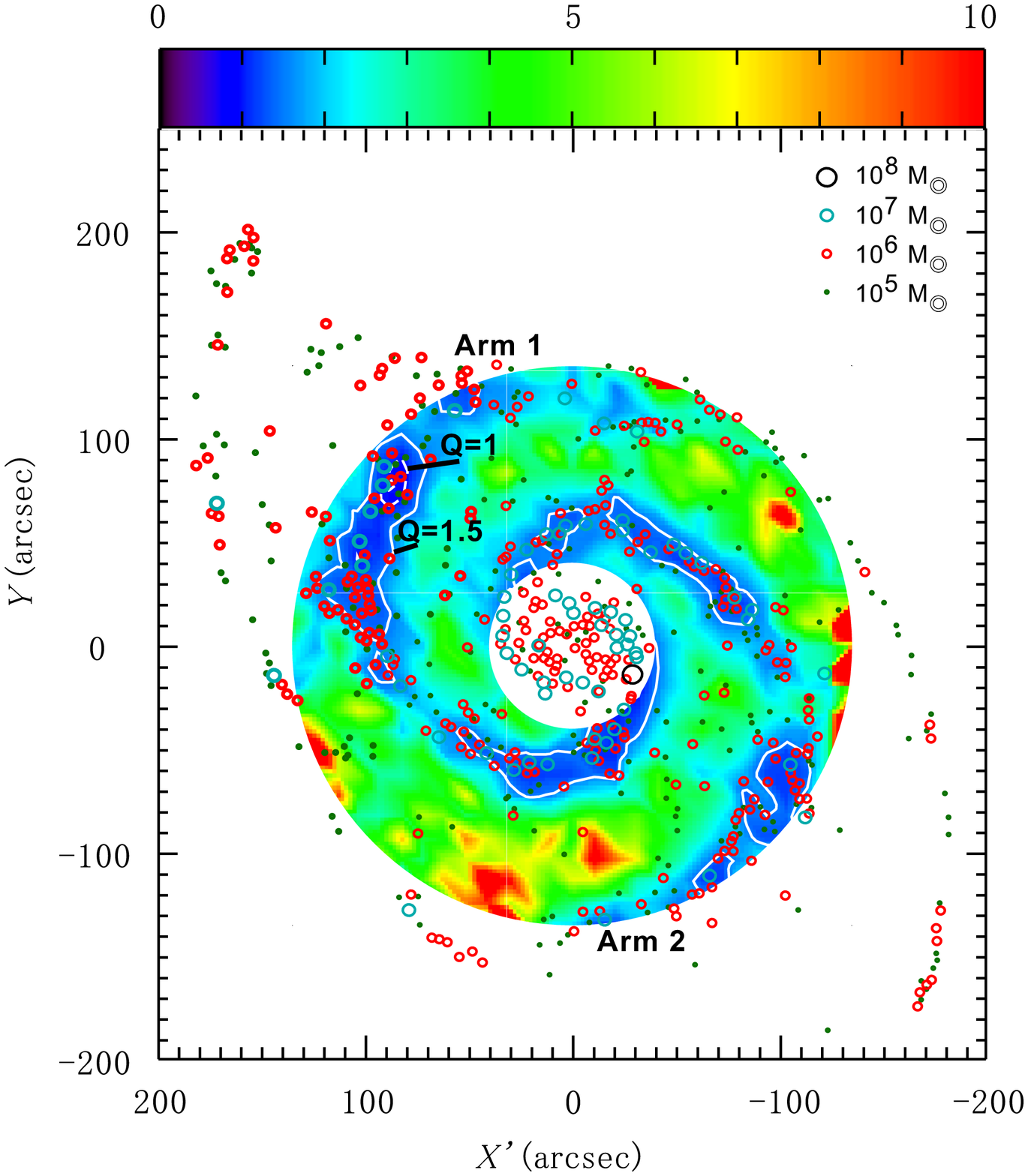}
   \end{flushleft}
   \end{minipage}
 \begin{minipage}{0.5\hsize}
\par
 \begin{flushleft}
  {\textbf{(b)}}
 \FigureFile(80mm,50mm) {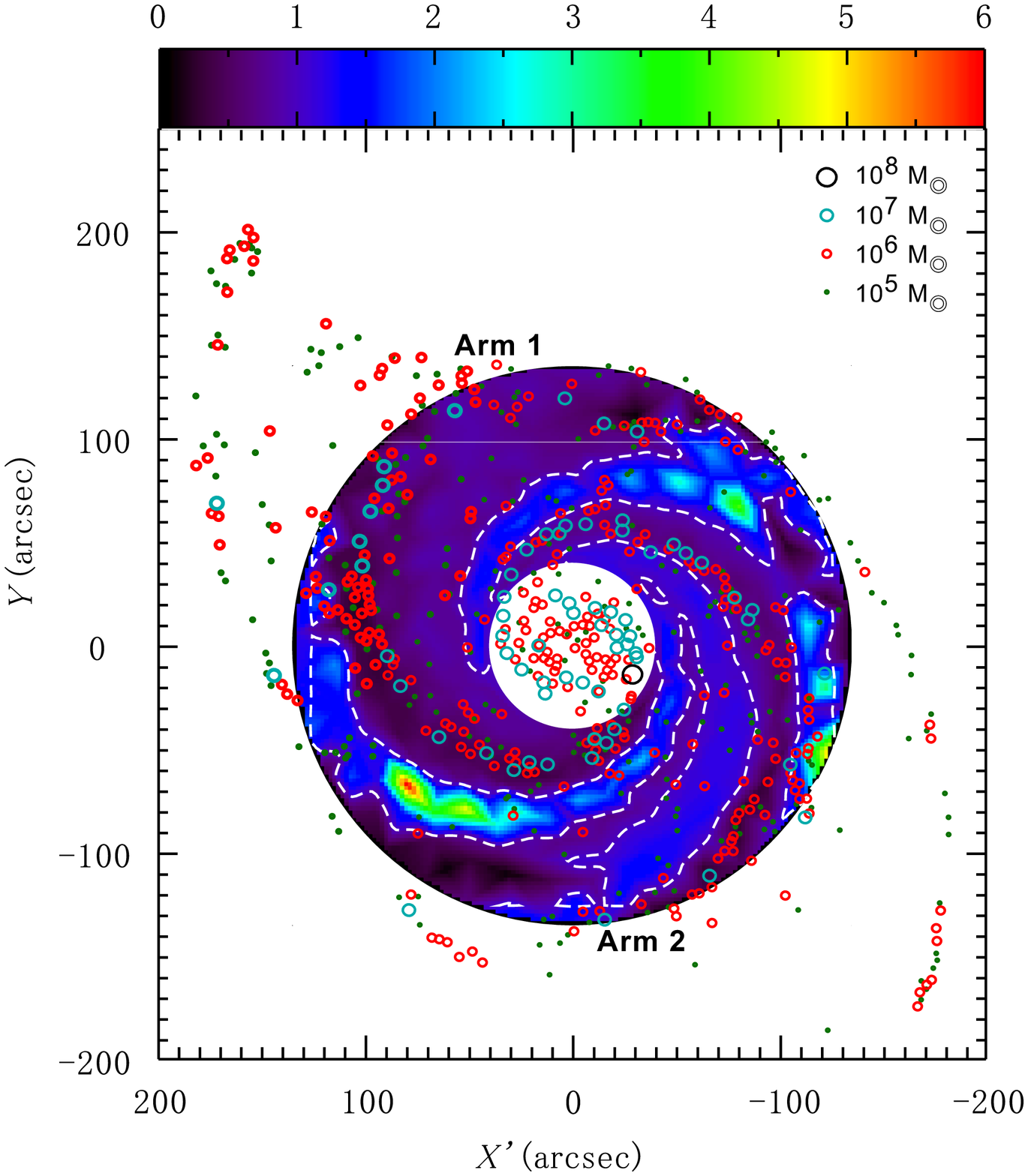}
 \end{flushleft}
 \end{minipage}
 \end{tabular}
\caption{The distributions of (a) $Q=\kappa v_{\rm s}/(\pi G\Sigma_{\rm gas})$ and (b) $Q^{\ast} = {\Sigma^{\rm shear}_{\rm crit}}/{\Sigma^{\rm grav}_{\rm crit}}$ (colour) overlaid on the distributions of GMAs ($\geq 10^7$~\MO) and GMCs ($\leq 10^6$~\MO) (open circles) shown by \citet{koda2009}.
The white solid lines in (a) correspond to $Q=1$ and 1.5, 
and the dashed lines in (b) $Q^{\ast}=1$.
}
 \label{fig:grav}
\end{center}
\end{figure}

Gravitational instability is one of the most probable mechanisms of GMA formation.
The criterion for gas disk stability can be defined by the Toomre's $Q$ parameter,
\begin{equation}
Q=\frac{\kappa v_{\rm s}}{\pi G \Sigma_{\rm gas}},
\end{equation}
where $\kappa$ is the epicyclic frequency (see section~\ref{sec:v_rot}), $G$ the gravitational constant, $v_{\rm s}$ the gas velocity dispersion, $\Sigma_{\rm gas}$ the gas surface density (\cite{toomre1964}, \cite{binney1987}). 
The critical surface density is described as 
\begin{equation}
\Sigma_{\rm crit}^{\rm grav}=\frac{\alpha\kappa v_{\rm s}}{\pi G}, 
\end{equation}
where $\alpha$ is a dimensionless constant.
While the constant of $\alpha=1$ has been applied in a thin stellar disk, 
$\alpha=0.67$ has been derived from the observed star formation thresholds in the disk galaxies \citep{kennicutt1989}.
When the gas surface density is larger than the critical density, the gaseous disk is gravitationally unstable to the formation of clumps such as GMAs.
The Toomre stability $Q$ in the disk of M51 is derived by adopting $v_{\rm s}\approx 10$~km~s$^{-1}$.
Figure \ref{fig:grav}(a) shows the distributions of $Q$, GMCs and GMAs.
The white contours show values of $Q=1$ and 1.5(=1/0.67).
The region of $Q<1.5$ corresponds to the spiral arms where  
GMAs (M$\geq 10^7$~\MO) and most of GMCs  (M$\leq10^6$~\MO) exist.

On the other hand,  
tidal shear motion can prevent the formation of GMAs by the gravitational instability 
when the shear force is stronger than the self-gravitational force.
Assuming a spherical and uniform density cloud, 
the cloud destruction by the shear overcomes the self gravity  
when the gas density is less than 
the critical shear density, $\rho^{\rm shear}_{\rm crit}$, 
presented by 
\begin{equation}
\rho^{\rm shear}_{\rm crit}=\frac{3 A (A-B)}{\pi G}, 
\end{equation}
where $B$ is Oort's $B$-constant defined by $B = -\Omega + A$  \citep{mihalas1968}. 
The critical shear surface density is simply defined as $\Sigma_{\rm crit}^{\rm shear} = 2h\rho_{\rm crit}^{\rm shear}$, where $h$ is the scale height of the gas disk.
The relative importance between the gravitational instability and the shear 
with respect to the evolution of GMAs can be determined from the ratio of the gravitational critical surface density, $\Sigma_{\rm crit}^{\rm grav}$, to the shear critical surface density,  $\Sigma_{\rm crit}^{\rm shear}$.
The ratio, $Q^{\ast}$, can be derived by 
\citet{kenny1993} as below, 
\begin{eqnarray}
Q^{\ast} &\equiv& \frac{\Sigma^{\rm shear}_{\rm crit}}{\Sigma^{\rm grav}_{\rm crit}}\nonumber\\
&=& 0.87\left(\frac{- A}{B}\right)^{1/2}, 
\end{eqnarray}
by adopting $h=v^2_{\rm z} / 2\pi G \Sigma_{\rm gas}$
as the scale height and assuming that the vertical velocity dispersion of $v_{\rm z}$ is comparable to the gas velocity dispersion $v_{\rm s}$ ($v_{\rm z}\approx v_{\rm s}$). 

The distributions of $Q^{\ast}$ and of GMCs and GMAs are displayed in figure~\ref{fig:grav}, where white dashed lines show a value of $Q^{\ast}=1$.
GMAs and most of GMCs clearly exist in the region of $Q^{\ast}<1$, although some GMCs in $Q^{\ast}\geqq1$.
These strongly suggest that GMAs can stably grow up  due to  the self-gravity and the accumulation of small clouds 
in the region such as the spiral arms where the shear critical density is less than the gravitational critical density, because  gravitational bind and instability are not interrupted by shear destruction. 
Some studies (e.g., \cite{kennicutt1989}, \cite{kenny1993}, \cite{luna2006}) have examined the relation between the tidal shear and formations of molecular clouds and stars.
They found that star formation is enhanced in the center region of galaxies and in the spiral arms
because the shear strength in those region is weak due to the solid-body rotation, 
and suggested that the tidal shear can control the star formation rate through cloud destruction.

To check the growth mechanism of molecular clouds, 
we compare the time scale of the GMA formation in the spiral arms and interarms
by gravitational instability and by collisional agglomeration 
with the GMA disruption time scale by the shear.
Here we representatively regard 
the arms ($Q^* <1$) and the interarms  ($Q^* >1$)
as the region of the spiral phase $\Psi= -$\timeform{30D}-$\timeform{30D}$, \timeform{330D}-\timeform{390D}  and $\Psi= $\timeform{110D}-$\timeform{170D}$, \timeform{450D}-\timeform{510D}, respectively.
The averaged crossing time scales in the arms and the interarms are $\tau_{\rm cross}\sim2\times$10$^7$~yr and $\sim1\times$10$^7$~yr, respectively, given by  
\begin{equation}
\tau_{\rm cross} =\frac{R-R_0} {V_{\rm r}}, 
\label{eq:vr}
\end{equation}
 where $V_{\rm r}$ ($=V \sin \phi$) is the radial velocity, 
typically $\sim -20$~km~s$^{-1}$ in the arms and $\sim6$~km~s$^{-1}$ in the interarms,
and ($R-R_0$) the radial travel distance of the molecular gas,
typically $\sim-350$~pc in the arms and $\sim50$~pc in the interarms (figure \ref{fig:orbit}).
The time scale of the gravitational instability can be estimated as follow \citep{larson1987}, 
\begin{eqnarray}
\tau_{\rm grav} &=& \frac{v_{\rm s}}{\pi G \Sigma_{\rm gas}}. 
\end{eqnarray}
In case that we use $v_{\rm s}=$10~km $\rm s^{-1}$ as typical velocity dispersion, 
we obtain the averaged time scale of  $\tau_{\rm grav}\sim2\times10^7$~yr in the arms ($\Sigma_{\rm gas} \approx 40$~\MO~$\rm pc^{-2}$; see figure~\ref{fig:velvec}) and $\sim4\times10^7$ yr in the interarms ($\Sigma_{\rm gas} \approx 20$~\MO~$\rm pc^{-2}$).
The shear time scale is simply derived as an inverse function of the gas speed relative to the pattern speed \citep{elmegreen1980};
\begin{equation}
 t_{\rm shear}=\frac{D}{R_{1} \left(\Omega(R_{1})-\Omega_{\rm p} \right) - R_{2} \left(\Omega(R_{2})-\Omega_{\rm p} \right)}\hspace{0.5em},
\end{equation}
where $D$ is the diameter of GMA, and
$\Omega_{\rm p}$ the pattern speed.
$\Omega(R_{1})$ and $\Omega(R_{2})$ are the angular velocities at the distances from the galactic center $R_{1}$ and $R_{2}$ ($R_{2} - R_{1}=D$), respectively.  
The adopted $D$ is the typical value of 500 pc (e.g., \cite{muraoka2009}) and 
the pattern speed $\Omega_{\rm p}$ is 38 km s$^{-1}$ kpc$^{-1}$ and 8 km s$^{-1}$ kpc$^{-1}$ at $R = 40'' $ - $110''$ and $110''$ - $140''$, respectively  (section~\ref{sec:v_rot}). 
The shear time scales in the arms and in the interarms are $\tau_{\rm shear}=4\times10^7$~yr and $1\times10^7$~yr, respectively.
Table~\ref{tab:timescale} summarizes these time scales.
In the arms, 
the gravitational instability time scale ($2\times10^7$~yr) is comparable to the crossing time scale ($2\times10^7$~yr)
and shorter than the shear time scale ($4\times10^7$~yr), 
while in the interarms, the gravitational instability time scale ($4\times10^7$~yr) is longer than the crossing ($1\times10^7$~yr) and the shear time scale ($1\times10^7$~yr). 
This means that GMAs can be stably formed by gravitational instability in the spiral arms.
The collisional time scale is given by
\begin{eqnarray}
\tau_{\rm coll} &=& \frac{1}{n_{\rm c} \sigma_{\rm coll} v_{\rm s}}, 
\end{eqnarray}
where $n_{\rm c}$ is the number density of GMC and GMA 
and $\sigma_{\rm coll}$ the  collisional cross section.
$n_{\rm c}$ is given by 
\begin{eqnarray}
n_{\rm c} &=& \frac{N}{2 S h},
\end{eqnarray}
where $S$ is the area of the arms 
and the interarms,
$h(=v^2_{\rm s}/2\pi G \Sigma_{\rm gas})$ the scale hight of the gas disk, 
and $N (=N_{\rm GMC}+N_{\rm GMA})$ the sum of the number of GMC~($N_{\rm GMC}$) and GMA~($N_{\rm GMA}$) counted by \citet{koda2009}.  
$N_{\rm GMC}$ and $N_{\rm GMA}$ are 83 and 22 in the arms, 
and 26 and 0 in the interarms, respectively.
The used area of the arms and the interarms are $S\approx6$~kpc$^{2}$,
for the range of the radius of $40''\leq R\leq 110''$ and the spiral phase width of $\Delta \Psi = \timeform{60D}$ (actual angular width of $\Delta\Psi/2=\timeform{30D}$; see equation (\ref{eq:spiral_phase})).
Adopting $v_{\rm s} =10$~km s$^{-1}$ 
and $\Sigma_{\rm gas} \approx 40$ and $20$~\MO~pc$^{-2}$
in the arms and interarms, respectively,  
the averaged number densities are $n_c\approx60$~kpc$^{-3}$ in the arms and $\approx3$~kpc$^{-3}$ in the interarms.
The effective radius $R_{\rm e}(=(N_{\rm GMC}D_{\rm GMC}+N_{\rm GMA}D_{\rm GMA})/N)$ of the collisional cross section ($\sigma_{\rm coll}=\pi R_{\rm e}^2$) is 144~pc in the arms and 50~pc in the interarms, 
assuming the diameter of GMC and GMA to be $D_{\rm GMC}=50$~pc (e.g., \cite{sanders1985}) and $D_{\rm GMA}=500$~pc (e.g., \cite{muraoka2009}), respectively.
The time scales of collisions between molecular clouds
are $\tau_{\rm coll}\sim2\times$10$^7$~yr in the arms and $\sim4\times$10$^9$~yr in the interarms for $v_{\rm s}=10$~km~s$^{-1}$ (table \ref{tab:timescale}).
Here we note that these collisional time scales are the upper limits,  
because we use the number counts of GMC and GMA in \citet{koda2009} 
in which the angular resolution was $4''$, corresponding to 186~pc at the distance of M51.
In order to discuss the time scale of collisions between GMCs with the size of a few 10~pc (e.g., \cite{sanders1985}) each other, 
observations with the higher angular resolution are required. 
These results however suggest that GMAs can be effectively grown up by collisional agglomeration and the gravitational instability in the spiral arms, 
while it is  difficult for GMA to be newly formed in the interarms.

\bigskip
In this paper we discussed the global influence of the galactic dynamics such as the spiral arm streaming motion and the shear on the evolution of GMC and GMA.
However, to discuss about fragmentation of GMA to GMC by shear and GMA formation by collisions between GMCs 
in more detail,  
the observations with higher spatial resolution enough to identify GMCs and without missing flux of extended gas are essential (e.g., ALMA).

\begin{table}
	\begin{center}
	\caption{Comparison of time scales}
		\begin{tabular}{ccc}
		\hline
		\hline
		(yr)							&Arm					&Interarm\\
		\hline
	$\tau_{\rm cross}$				&2$\times10^7$			&1$\times10^7$\\
	$\tau_{\rm grav}$				&$2\times10^7$			&$4\times10^7$\\
	$\tau_{\rm coll}$				&$<2\times10^7$			&$<4\times10^9$\\
	$\tau_{\rm shear}$($D=500$~pc)	&$4\times10^7$			&$1\times10^7$\\
		\hline
		\hline
		\end{tabular}
		\label{tab:timescale}
	\end{center}
\end{table}%


\clearpage
\section{Conclusions}
\label{chap:con}
The $^{12}$CO $(J=1$$-$$0)$ line emission in the region of about $\rm9' \times 10'$ (25~kpc$\times$28~kpc) of the spiral galaxy M51 was mapped with the NRO 45-m telescope.
We investigated the kinematics and distribution of the molecular gas using the CO data.
The main conclusions are summarized as follows:
\begin{enumerate}
\item The position angle of the major axis of the galactic disk systematically varies from about $\theta_{\rm PA}=-\timeform{3D}$ to $-\timeform{15D}$ with increasing the radius from $R=40''$ to $140''$.  
The systematic variation of the position angle with radius could be due to 
warping of the galactic disk caused by the tidal effect of the companion galaxy NGC 5195 
rather than non-circular motion of the molecular gas.
The systemic velocity is $V_{\rm sys~(LSR)}=\rm469\pm4$~km~s$^{-1}$.
The inclination angle is $i=\timeform{22D}\pm\timeform{3D}$, which was derived from the baryonic  Tully - Fisher relation.
The outer disk traced by HI is more inclined than the inner disk.

\item The radial distribution of the surface density of the molecular gas shows an exponential decrease of $\sigma$(H$_2$) $= 202$ $\exp [-R / 2$~kpc]~\MO~pc$^{-2}$ at $1 \leq R \leq 4$~kpc where the rotation curve shows the differential rotation, 
suggesting the inflow of the molecular gas due to viscosity of the gas. 
At $R < 1$ kpc of the rigid rotation, $\sigma$(H$_2$) dips below the value expected from the exponential shape. 
At $4 < R < 10$ kpc, $\sigma$(H$_2$) shows a local maximum at $R \approx$ 6 kpc 
due to the kinks or fractures of the spiral arms, 
caused by an overall pattern of non-circular motions associated with the interaction with the companion galaxy NGC 5195. 
The increase of $\sigma({\rm H_2})$ at $R \approx 12$ kpc is due to the molecular gas in NGC 5195

\item Two molecular arms are fitted by the logarithmic spirals with a pitch angle of $p=\timeform{19D}\pm\timeform{1D}$ at $40'' \leq R\leq 140''$.
The location of OILR and CR is  $R = 39''$ and $115''$, respectively, for the pattern speed of $\Omega_{\rm p}=38$~km~s$^{-1}$~kpc$^{-1}$.

\item The velocity components parallel and perpendicular to the spiral arms were derived from the distribution of the line-of-sight velocity of the molecular gas in each spiral phase.
In the inner region ($R = 40''$ - $110''$) of the M51 disk, the variations of the velocities and the density distribution of the molecular gas with respect to the spiral potential are qualitatively in good agreement with the density-wave theory in the particle-system model. The streamline of the molecular gas derived from the velocity vectors shows an oval as predicted by the density wave theory.
In the outer region ($R = 110''$ - $140''$), however, 
the orbit deviates from the shape expected from the theory, 
because the motion of the molecular gas is affected by the tidal interaction with NGC 5195.

\item The distributions of GMAs and the shear strength due to the differential rotation show the anticorrelation each other.
GMAs exist only in the area of the weak shear strength and on the upstream side of the high shear strength. 
The strong shear motion disrupts GMAs into the smaller clouds of GMCs which are ejected into the interarms. 
These indicate the shear motion is an important factor in evolution of  GMAs. 

\item GMAs and most of GMCs exist in the regions where the shear critical surface density is smaller than  the gravitational critical surface density. 
GMAs in the arms can be stably formed by self-gravity and grow quickly by the collisional agglomeration  of small clouds without being destroyed by shear motion, 
but GMAs can not be newly formed in the interarms.
\end{enumerate}

\bigskip
We thank M. Horie, K. Hagiwara, K. Mamyoda, S. Ishii, T. Masuda and N. Miyagawa for helping with the CO observations using Nobeyama 45-m telescope, 
and D. Salak for helpful comments.
Thoughtful comments by Dr. L. Bronfman are gratefully acknoledgement. 
We thank the referee, J. Kenney, for valuable comments on the manuscript.
The Nobeyama Radio Observatory is a branch of the National Astronomical Observatory of Japan, National Institutes of Natural Sciences.
This research has made use of the NASA/IPAC Extragalactic Database (NED) which is operated by the Jet Propulsion Laboratory, California Institute of Technology, under contract with the National Aeronautics and Space Administration.
We acknowledge the usage of the HyperLeda database\footnotemark[$\dagger$].
\footnotetext[$\dagger$]{http://leda.univ-lyon1.fr}

\end{document}